\documentclass[longauth]{aa} 

\usepackage{graphicx}
\usepackage{txfonts}
\usepackage{hyperref}
\AtBeginDocument{%
  \pdfstringdefDisableCommands{%
    \def\&{and }%
  }%
}
\usepackage{amsmath}
\usepackage{multirow}
\usepackage{longtable}

\newcommand{\kms}{km\,s$^{-1}$}
\newcommand{\vsini}{$v\sin i$}

\newcommand{\vrad}{$v_{\rm rad}$}

\newcommand{\msun}{M$_{\odot}$}

\newcommand{\bl}{$B_{\ell}$}

\newcommand{\ddeg}{$^{\circ}$}
\newcommand{\te}{$T_{\rm eff}$}
\newcommand{\Bpol}{$B_{\mathrm{pol}}$}
\newcommand{\logg}{$\log{g}$}

\usepackage[dvipsnames]{xcolor}
\usepackage{soul}
\soulregister\cite7
\soulregister\citet7
\soulregister\citep7
\soulregister\ref7
\soulregister\pageref7
\setul{0.3ex}{0.3ex}

\begin{document}

   \title{Magnetic field measurements in a sample of Class~I and flat-spectrum protostars observed with SPIRou}

   \author{L.~Drouglazet
          \inst{1}\thanks{Based on observations obtained at the Canada-France-Hawaii Telescope (CFHT) which is operated by the National Research Council of Canada, the Institut National des Sciences de l'Univers of the Centre National de la Recherche Scientifique of France, and the University of Hawaii.}
          \and
          E.~Alecian\inst{1}
          \and
          A.~Sousa\inst{1}
          \and
          P.I.~Cristofari\inst{2}
          \and
          E.~Artigau\inst{3,4}
          \and
          J.~Bouvier\inst{1}
          \and
          A.~Carmona\inst{1,5}
          \and
          N.J.~Cook\inst{3}
          \and
          C.~Dougados\inst{1}
          \and
          G.~Duchêne\inst{1,6}
          \and
          C.P.~Folsom\inst{7}
          \and
          H.~Nowacki\inst{8}
          \and
          K.~Perraut\inst{1}
          \and
          S.H.P.~Alencar\inst{9}
          \and
          L.~Amard\inst{10}
          \and
          M.~Audard\inst{10}
          \and
          S.~Cabrit\inst{11,1}
          \and
          J.-F.~Donati\inst{5}
          \and
          K.~Grankin\inst{12}
          \and
          N.~Grosso\inst{13}
          \and
          O.~Kochukhov\inst{14}
          \and
          \'A.~K\'osp\'al\inst{15,16,17}
          \and
          V.J.M.~Le~Gouellec\inst{18,19}
          \and
          L.~Manchon\inst{20}
          \and
          G.~Pantolmos\inst{21}
          \and
          P.~Petit\inst{5}
          \and
          L.~Petitdemange\inst{22}
          \and
          R.~Devaraj\inst{23}
          \and
          H.~Shang\inst{24}
          \and
          M.~Takami\inst{24}
          }

   \institute{Univ. Grenoble Alpes, CNRS, IPAG, 38000 Grenoble, France,    
   \email{lisa.drouglazet@univ-grenoble-alpes.fr}
         \and
         Leiden Observatory, Leiden University, PO Box 9513, 2300 RA Leiden, The Netherlands
         \and
             Institut Trottier de recherche sur les exoplanètes, Département de Physique, Université de Montréal, Montréal, Québec, Canada
        \and
             Observatoire du Mont-Mégantic, Québec, Canada
        \and
            Université de Toulouse, CNRS, IRAP, 14 avenue Belin, 31400 Toulouse, France
        \and
            Department of Astronomy, University of California, Berkeley CA, 94720, USA
        \and
        Tartu Observatory, University of Tartu, Observatooriumi 1, 61602, Toravere, Estonia
        \and
            Université Côte d'Azur, Observatoire de la Côte d'Azur, CNRS, Laboratoire Lagrange, France
         \and
         Departamento de Fisica – ICEx – UFMG, Av. Antônio Carlos 6627, 30270-901 Belo Horizonte, MG, Brazil
         \and
             Department of Astronomy, University of Geneva, Chemin Pegasi 51, 1290 Versoix, Switzerland
        \and
            Observatoire de Paris, PSL University, Sorbonne Université, CNRS, LERMA, F-75014, Paris, France
        \and
            Crimean Astrophysical Observatory, 298409, Nauchny, Republic of Crimea
         \and
            Aix Marseille Univ, CNRS, CNES, LAM, Marseille, France
         \and
             Department of Physics and Astronomy, Uppsala University, Box 516, SE-751 20 Uppsala, Sweden
        \and
        Konkoly Observatory, HUN-REN Research Centre for Astronomy and Earth Sciences, MTA Centre of Excellence, Konkoly-Thege Mikl\'os \'ut 15-17, 1121 Budapest, Hungary
        \and
        Institute of Physics and Astronomy, ELTE E\"otv\"os Lor\'and University, P\'azm\'any P\'eter s\'et\'any 1/A, 1117 Budapest, Hungary
        \and
        Max-Planck-Insitut f\"ur Astronomie, K\"onigstuhl 17, 69117 Heidelberg, Germany
        \and
        Institut de Cienci\`es de l’Espai (ICE-CSIC), Campus UAB, Carrer de Can Magrans S/N, E-08193 Cerdanyola del Vall\`es, Spain
        \and
        Institut d’Estudis Espacials de Catalunya (IEEC), c/ Gran Capitá, 2-4, 08034 Barcelona, Spain
        \and
            LESIA, Observatoire de Paris, PSL Research University, CNRS, Université Pierre et Marie Curie, Université Denis Diderot, 92195 Meudon, France
        \and
            Department of Physics, National and Kapodistrian University of Athens, University Campus, Zografos GR-157 84 Athens, Greece
        \and
            LERMA, Observatoire de Paris, PSL Research University, CNRS, Sorbonne Université, Paris, France
        \and
            School of Cosmic Physics, Dublin Institute for Advanced Studies, 31 Fitzwilliam Place, Dublin 2, Ireland
        \and
        Institute of Astronomy and Astrophysics, Academia Sinica, Taipei 106319, Taiwan
}

   \date{Received November 25, 2025; accepted March 12, 2026}

 
  \abstract
   {Magnetic fields play a crucial role throughout stellar evolution in regulating angular momentum, channelling accretion, and launching jets and outflows. While the magnetic properties of classical T Tauri stars (CTTSs) have been extensively characterised, those of their progenitors, Class~I and flat-spectrum (FS) protostars, remain poorly constrained due to their embedded nature, which provides an observational challenge.}
   {Our aim was to detect and characterise the large-scale magnetic fields in a sample of Class~I and FS protostars. These young stars are expected to host strong magnetic fields generated by dynamo processes in their largely convective interiors.} 
   {We have used SPIRou, a high-resolution spectropolarimeter working in the near-infrared domain, to analyse the polarised light of Class~I and FS protostars. We used the least-squares deconvolution (LSD) technique to perform the magnetic analysis and measure the longitudinal magnetic fields from circularly polarised Stokes $V$ profiles.
}
   {We report new large-scale magnetic field detections in five FS protostars. Including the previous detection of the large-scale magnetic-field  in the V347 Aur FS-protostar, 40\% of our final sample of 15 protostars is confirmed to be magnetic. These magnetic stars show clear Zeeman signatures, with longitudinal field strengths ranging from $\sim$80 to $\sim$200 G in absolute value. The remaining stars exhibit no detectable Stokes~$V$ signature, but the estimated upper limits on a hidden dipolar field range from 500~G to more than 5~kG. For stars in which no magnetic fields are detected, it is still conceivable that a magnetic field exists, but is  intrinsically weak, highly complex and dominated by small-scale structures, or  cancelled out in integrated spectropolarimetric signals due to opposing polarities.} 
   {We show that Class~I and FS protostars can host large-scale magnetic fields with strengths that may be weaker than in more evolved CTTSs. This supports the idea that magnetic processes are already active during the main accretion phase and may influence star--disk interactions from the earliest stages.}


   \maketitle
%
\section{Introduction}
    
   Star formation occurs in dense cores within molecular clouds. These clouds are magnetised, and the cores, called pre-stellar cores \citep{Ward-Thompson1994}, collapse under gravity, eventually giving birth to protostars. During this early evolution, magnetic fields are thought to play a significant role in several processes, such as accretion, ejection, and interaction with the circumstellar environment \citep{Lada1987}. Through their role in these processes, which significantly impact stellar rotation and evolution, magnetic fields shape the close environment of the protostar \citep{Bouvier2007,Vaytet2018}.

    Regulation of angular momentum is made through magnetospheric accretion and ejection processes during the pre-main sequence (PMS) phase \citep{Bouvier2014, Lee2017}. In classical T Tauri stars (CTTSs), the presence of large-scale magnetic fields is well established \citep{Donati2011}. These fields, typically of kilogauss strength, are responsible for truncating the inner disk and channelling the accreted material onto the stellar surface \citep{Bouvier2007, Bouvier2014, Hartmann2016}. The magnetic properties of CTTSs (between 0.5 and 1.3 $M_{\odot}$) are characterised by a dominant poloidal field, with an average strength between 1 and 2 kilogauss \citep{Zaire2024}.

   While magnetism and its impact on the star--disk interaction have been intensely explored during the PMS phase \citep{Romanova2015b}, our current knowledge of the magnetic properties and of the star--disk interaction in earlier phases remains limited. During those phases, the protostar experiences intense accretion, often through bursts, and grows in mass until it approaches its final mass \citep{Baraffe2009, Fisher2023}. Such major internal structural changes may have a strong impact on the stellar magnetic field, which in turn may influence the connection between the star and its accretion disk \citep{Kunitomo2017}. The progenitors of the CTTSs (Class~II) are the Class~I protostars. These objects are still embedded within their envelopes of gas and dust \citep{Lada1987}. The protostars transiting from Class~I to Class~II are observed as flat-spectrum (FS) sources \citep{Greene1994}. Due to their dense and obscuring environments, the direct observation of Class~I protostars and the detection of photospheric features is challenging \citep{Connelley2010, Doppmann2005}, making it difficult to study their surfaces, and therefore their magnetic fields. 
   
   For several reasons we expect Class~I protostars to be strongly magnetised, as in CTTSs. They have evolved enough to set off convection in their full interior \citep{Palla1998,Baraffe2017}, and are thus   capable of hosting strong magnetic fields generated by dynamo processes. In other types of fully convective stars, such as the M dwarfs on the main sequence, strong dynamo magnetic fields are routinely observed \citep{Donati2008,Morin2010,Donati2023}. Signs of magnetospheric accretion, as in the CTTSs, have been detected in Class~I sources \citep{Doppmann2005, Fiorellino2023}, suggesting the presence of strong large-scale magnetic fields at the surface of the embedded star. 
Powerful flares have been detected in X-ray observations of Class~0 and Class~I protostars \citep{Grosso1997, Grosso2020, Montmerle2000}, similar to those present in the Sun or more evolved stars, but much more energetic \citep{Feigelson2007}. These emissions are clues of intense magnetic activity. These studies therefore suggest that strong magnetic fields may already be in place during the Class~I phase and may regulate mass inflow and angular momentum loss early on.

A recent study by \citet{Flores2024} has investigated a sample of 42 Class~I and FS protostars, and measured their total surface magnetic field from the Zeeman broadening of spectral lines. They found a median magnetic field strength of 1.9\,kG, which is consistent with the value found in Class~II sources and with earlier studies of Class~I protostars \citep{Johns-Krull2009}, indicating that Class~I and FS sources are indeed strongly magnetic. However, this study cannot inform us on the large-scale components of their magnetic fields, which interact with the disk.
To understand whether magnetospheric accretion and ejection processes are in place at that age, and are potentially at the origin of the jets observed in almost all Class~I sources, we need to characterise their large-scale magnetic field.

Furthermore, dynamo processes are not well understood, especially in fully convective stars. The imprint of the dynamo processes in stellar evolution is still an open question \citep{Moss2003,Donati2008}. In particular, simulations have shown that the initial conditions at the launch of the dynamo may impact the future properties of dynamo magnetic fields, and may explain the bistability of magnetic fields observed in fully convective M dwarfs \citep{Gastine2012,Gastine2013,Raynaud2015}. While the magnetic energy produced by dynamo processes is an important characteristic to constrain the models, another factor is the ratio of poloidal to toroidal fields, which is only accessible by measuring the magnetic geometry.

The topology and strength of the large-scale magnetic field can be obtained with the Zeeman Doppler imaging (ZDI) technique \citep{Semel1989}. This technique uses time series of circularly polarised and intensity spectra well sampled over the rotation period of the star using a high spectral resolution spectropolarimeter. By assuming that the magnetic field components can be modelled as spherical harmonics, and by using an inversion technique, ZDI provides us with the maps of the radial, azimuthal, and meridional components of the field \citep{Donati2006, Kochukhov2016}. This technique has been largely employed to map the magnetic field of various types of stars, including CTTSs \citep[e.g.][]{Donati2009,Zaire2024}.To date, ZDI has not been largely performed in Class~I and FS protostars.

 Recent advances in instrumentation have opened new possibilities for investigating these elusive protostars. Among these instruments is the SPectropolarimètre InfraRouge (SPIRou), a spectropolarimeter installed on the 3.6-meter Canada-France-Hawaii Telescope (CFHT). SPIRou \citep{Donati2020} operates in the near-infrared (nIR) domain, which allows us to pierce through their dusty envelope to probe the photosphere of Class~I protostars. SPIRou’s high-resolution, high-sensitivity measurements in the $YJHK$ bands are particularly well suited to study Class~I protostars, which are largely inaccessible to optical spectropolarimeters. 

To investigate the magnetic properties of Class I protostars, and their accretion--ejection mechanisms, we have built the \emph{Protostellar Magnetism: Heritage and Evolution}\footnote{https://promethee-anr.github.io/} (PROMETHEE) project. This project will determine for the first time the large-scale magnetic properties of protostars, and the structure of their magnetospheres, and will build unique MHD models of magnetic protostars from their birth to the T Tauri phase. This paper is the first in a series that describes the first SPIRou analysis of a sample of Class~I and FS protostars.
These data have also been used to study the accretion and ejection diagnostics available in the SPIRou spectra (e.g. HeI 1083 nm, Pa$\beta$ or Br$\gamma$ emission lines, or continuum veiling). These results will be submitted in an associated paper (Sousa et al., in prep.). In the meantime, within the PROMETHEE project, we are developing dynamo models adapted to young fully convective protostars, and interacting with their disk (Guseva et al., in prep.). In order to understand the magnetic and stellar properties at the start of the protostellar phase, and how they can subsequently affect the dynamo processes, we are also developing models of molecular cloud collapses until the development of the second Larson core \citep[][, Ahmad et al., in prep.]{Ahmad2025a,Ahmad2025b}

Before performing a monitoring for each source in our sample of Class~I and FS protostars, to map their magnetic fields, we performed a snapshot survey with SPIRou. Our aim was to first detect their Zeeman signature in circularly polarised (Stokes $V$) spectra. In future work, we will monitor with SPIRou only those with a Zeeman detection to perform ZDI.

In this study, we present the analysis of our snapshot sample. It is structured as follows. We describe the observations, data reduction, and spectral and polarimetric analyses in Sect.~\ref{sec:obs}. Our results are presented in Sect.~\ref{sec:analyse} and discussed in Sect.~\ref{sec:discussion}. We summarise our conclusions in Sect.~\ref{sec:ccl}.


\section{Observations}
\label{sec:obs}

\subsection{Sample selection} 

Our goal is to characterise the magnetic fields of a sample of Class~I and FS objects. To this aim, we need to detect the circular polarisation signal induced by the Zeeman effect in the photospheric lines. To build our sample, we first listed the objects in which photospheric lines had been detected in previous surveys using high-resolution optical or nIR spectrographs \citep{White2004, Doppmann2005, Evans2009, Connelley2010, Rebull2010, Almeida2012, Dunham2015}.
Then we first kept the sources north to a declination of -40\ddeg, to ensure that they were observable from the CFHT. In a second step, we selected sources brighter than 10 magnitudes in the H band so that the total exposure time needed to detect a typical dipolar magnetic field of $\sim$0.5\;kilogauss is within 1.5 h, assuming a typical veiling in the H band of about 2 \citep{Doppmann2005}. In the star-forming regions that we explored, Class~I and FS sources display H band magnitudes typically ranging from ~8 to ~12, depending on how deeply embedded they are or how inclined their bipolar cavity is with respect to our line of sight. As a consequence, the magnitude criterion may introduce a bias towards  the less embedded objects, i.e. more evolved Class~I objects, or towards low inclination. Two of our targets have magnitudes H greater than 10. We had an opportunity to obtain additional observations, so we included them in our sample.

Our initial sample is composed of 18 Class~I (6) and FS (12) objects located in different nearby star-forming regions: Perseus, Taurus, rho Ophiuchi, Serpens, and Corona Australis. When known, the spectral types of the observed sources range from G7 to M6, with the majority falling between K1 and M4. The projected rotational velocities (\( v \sin i \)), when known, range from 5 to 52 km/s. These properties are representative of the typical characteristics observed in embedded protostellar objects \citep{White2004, Doppmann2005}.

Most of the data have been obtained with PI programs allocated over different semesters (PI: E. Alecian). We have also included data from a few Class~I sources (V347 Aur, HL Tau, the first observation of V512 Per 2020, GY92 214 and GY92 378) obtained during the SPIRou Legacy Survey (SLS) large program (PI: J.-F. Donati, \citealt{Donati2020}).

The complete log of the observations is provided in Table \ref{tab:log}. The data of V347~Aur, which have already been published in \citet{Donati2024}, do not appear in that table.

\subsection{SPIRou observations and data reduction}

The primary goal of SPIRou is to detect and characterise exoplanets around M dwarfs, but also to study magnetic fields in low-mass mature and PMS \citep{Donati2020}. SPIRou covers a spectral range from 950 nm to 2\,500 nm, with a resolution of 70\,000 and a velocity bin per pixel of 2.3 \kms \citep{Donati2020}. It consists of an achromatic polarimeter, which splits light into two beams associated with orthogonal states of the selected polarisation, transmitted via optical fibres to an ultra-stable thermally controlled cryogenic spectrograph. The spectrograph disperses light using an echelle grating and a double-pass cross-dispersing prism train, recording the spectra on an H4RG detector.

The polarimeter consists of Fresnel rhombs and a Wollaston prism, enabling simultaneous measurements of both orthogonal states of the selected polarisation (circular polarisation in the present case). The operational process includes stabilising the stellar image on the entrance aperture, injecting it into optical fibres, and using calibration sources (a Fabry-Pérot interferometer, in our case) for precise wavelength and radial velocity calibration.

The camera captures raw spectral data, with echelle orders spread across a 4K × 4K pixel detector. These data include wavelength and light intensity information, enabling the extraction of parameters such as radial velocities and magnetic signatures of stars. The spectrum is divided into 49 orders, each of which requires individual normalisation to maintain the uniformity and precision of the analysed data. Each order comprises two spectra, corresponding to left and right circular polarisations. For each stellar observation, four sub-exposures are taken at successive time intervals with a 90\ddeg\ rotation of the polariser, which helps mitigate instrumental biases during the extraction of the polarisation signal. 

We used V0.7.292 version of the dedicated APERO\footnote{https://apero.exoplanets.ca/} pipeline \citep{Cook2022} to reduce the observations. 
Compared to version V0.7.255 described in \citet{Cook2022}, in the APERO versions there have been changes to the background subtraction. The thermal background spectrum is obtained by averaging the spectra acquired from telescope dark frames (i.e. with no sky scene injected). The thermal spectrum is then scaled to match the thermal emission of saturated absorption lines in the K band. An issue was noted, as persistence structures in the J band led to spurious subtraction, being mistaken for low-level thermal emission. Since one does not expect a measurable level of thermal emission below 1.9 µm (considering the rapid decline of a near-room-temperature blackbody bluewards of this wavelength), the thermal emission spectrum was set to zero below 1.9 µm in the newest versions of APERO. It should be noted that, in practise, the thermal background is only subtracted within the K band; thus, this zeroing of the thermal emission at the blue end does not affect the overall thermal background correction. Finally, all individual spectra were corrected for telluric absorption (including OH lines), as described in \citet{Cook2022}.

\subsection{Continuum normalisation of individual spectra}

To analyse the magnetic signals within the photospheric lines, a normalisation of the spectra to their continuum is needed. Automatic continuum normalisation can be performed during the extraction of the polarised spectra. However, the spectra of Class~I sources have several variable emission lines that need to be taken into account on a case-by-case basis. While normalisation tools already exist, they lack specific options we aimed to incorporate (necessary for emission line spectra such as those of Class~I sources). We developed a normalisation tool for SPIRou data files called \texttt{Fantasio}.\footnote{https://github.com/Lisa2626/Fantasio} The code aims to reproduce the principles of the IRAF normalisation tool called continuum.
The selection of the continuum points is performed by removing non-continuum points using a sigma-clipping procedure over several iterations. The normalisation is performed order by order when a 2D SPIRou spectrum is given as input. 
The developed interface allows users to modify normalisation parameters interactively for each order, including selecting the polynomial order of the spline and the number of spline interpolations. The sigma-clipping parameters, such as the upper and lower thresholds, and the number of iterations, can also be adjusted to optimise the fit. We allow different threshold values for points above and below the continuum to provide flexibility when both photospheric and emission lines are present within one order. Another interactive feature enables the removal of photospheric and emission lines to exclude them from the fitting process by selecting the regions to reject. This is particularly useful when dealing with spectra containing numerous strong emission lines, as in Class~I sources. Normalisation can be performed manually by selecting parameters for each order, or it can be run automatically for multiple observations using pre-defined parameters. 

\begin{figure*}[h]
\centering
\includegraphics[width=0.9\linewidth]{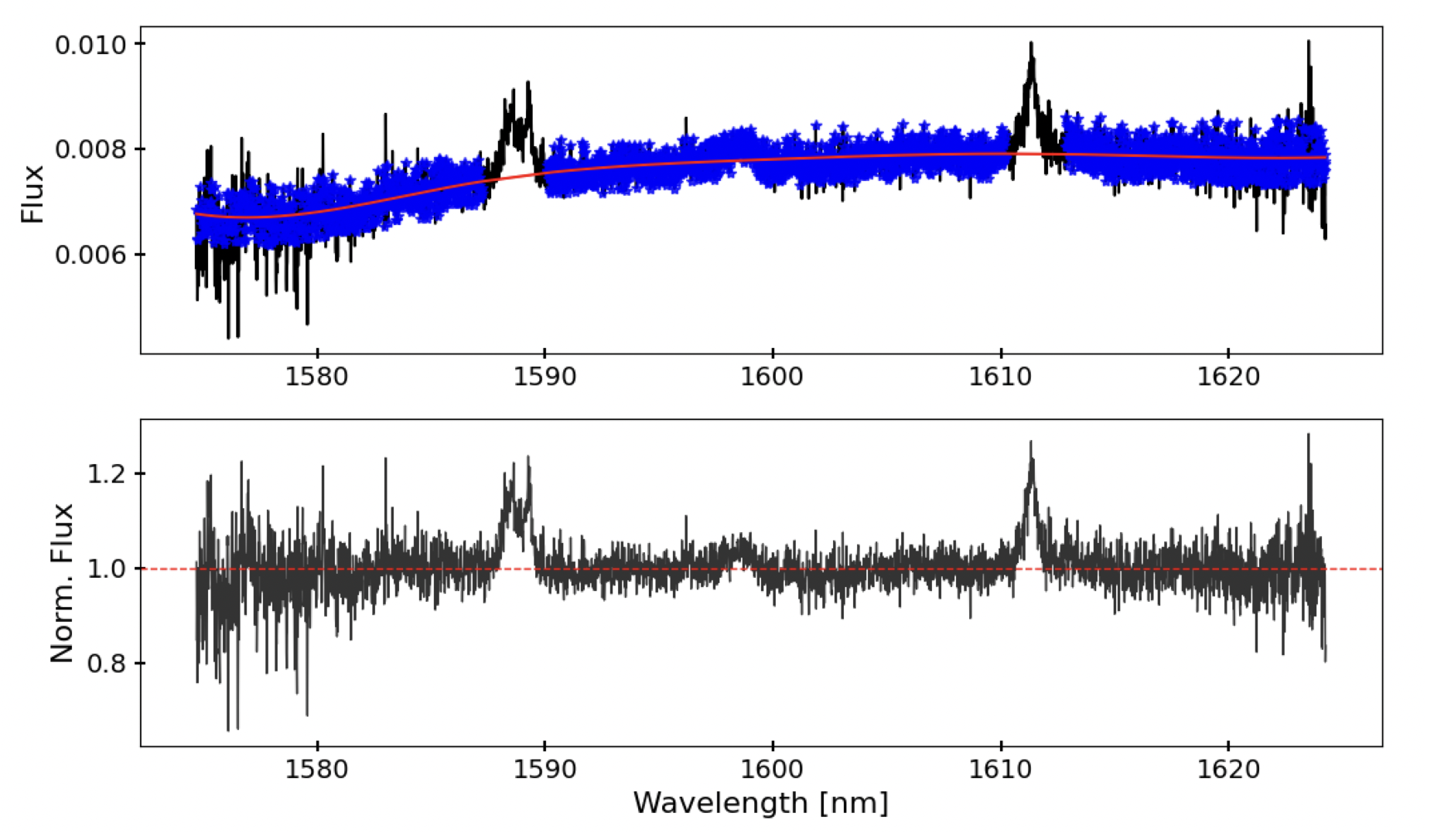}
\caption{\label{Norm}{\it Upper panel}: Order 32 of the spectrum of VV\,CrA\,SW without normalisation (solid black  line). In blue are the data points selected to fit the continuum using a sigma-clipping of 3 and 2 iterations. The spline fit of order 3 is plotted in red with a number of knots of 2. The two regions with emission lines were manually removed using the interactive tool. {\it Lower panel}: Same spectrum after normalisation. The dashed red line shows the continuum normalised to 1.}
\end{figure*}

We performed a manual normalisation on both A- and B-spectra, corresponding to the two orthogonal polarisations, for one observation of each star. We then applied the automatic normalisation to all other observations of the same star, using the same fitting parameters per order as those chosen for the fit of the first observation. In general, the order of the spline fit ranges from 1 to 3, as the number of knots of the spline. The sigma clipping can vary from 1 to 3 sigma, and 3 iterations are used. An example of the normalisation process is shown in Fig. \ref{Norm}.

\subsection{Computation of the polarised spectra}

The \texttt{spirou-polarimetry} package of \citet{Martioli2020} was used to extract the polarised spectra from our normalised individual intensity spectra in both polarimetric channels. 
The Stokes $V$ spectrum is obtained by using the ratio method, when the channels A and B spectra are recombined \citep{Martioli2020}. The unpolarised intensity spectrum Stokes $I$ is obtained by adding all channels and sub-exposure spectra. Finally, the null polarisation spectrum ($N$) is obtained by combining the A and B channels of the four sub-exposures, in a way that cancels the input circular polarisation. It is used to diagnose potential spurious polarised signal (due to instrumental, data reduction issues, or stellar pulsations).

The polarimetry pipeline computes the Stokes spectra by identifying all sequences consisting of four sub-exposures, ensuring that the polarimeter is positioned at the correct angle for each, and calculating the corresponding polarimetric spectra for each sequence, including their associated uncertainties, as described in \citet{Bagnulo2009}. The Barycentric Earth Radial Velocity (BERV) shift is corrected before calculating the polarimetric spectra. This pipeline was originally developed to use non-blazed corrected, and non-continuum normalised individual spectra. We have therefore adapted the code to allow the use of our normalised individual spectra. This has also impacted the computation of the uncertainties, which needed to be divided by the blaze and normalised by the continuum. All spectra are extracted with respect to the intensity continuum ($I_{\rm c}$), namely $I/I_{\rm c}$, $V/I_{\rm c}$, and $N/I_{\rm c}$. In the following, we adopt the simplified writing, $I$, $V$, and $N$ as their normalised value.

\section{Analysis and results}
\label{sec:analyse}

In this section we present the results of our analysis performed on the photospheric lines of the spectra, both in intensity and polarisation, with the final objective being to detect the Zeeman signature in the Stokes $V$ spectra, and to determine either the magnetic strength or an upper limit on the dipolar magnetic field when no magnetic field is detected. We detected photospheric lines in the spectra of all our sources, except for one: the FS protostar IRAS 03413+3202. Furthermore, the signal-to-noise ratio (S/N) values obtained for two Class~I sources with H magnitude greater than 10 (GY92\,214 and GY92\,378) are too low to constrain either their effective temperatures or their magnetic fields. In the following, we   therefore exclude those three objects. In addition, the dataset of V347 Aur has already been analysed and published in \citet{Donati2024}. The methods used in this paper are similar to those we have used on our data. We have therefore not re-done the analysis for this object, but we kept the source in our sample for the statistical analysis we performed and in the discussion. We present below the photospheric line analysis we performed on the 14 remaining sources.

\subsection{Effective temperature determination}
\label{sub:temp}

We employed two complementary approaches to estimate the effective temperatures (\te) of our targets. Our primary method consisted of fitting the observed spectra with synthetic spectra computed using the \texttt{MARCS} atmospheric models \citep{Gustafsson2008, Plez2012} and the spectrum-synthesis code \texttt{ZeeTurbo}, following the procedure described in \citet{Cristofari2023}. 
We assumed solar metallicity and a fixed microturbulent velocity of 1\,\kms\, and broadened with a Gaussian profile of full width at half maximum (FWHM) of 4.3 \kms\ to account for the instrumental width. In our fitting process, we fixed the macroturbulence to 0 \kms\ and allowed the projected rotational velocity (\vsini) and veiling in the H and K bands to vary. 
First, we also allowed the magnetic field to vary.
However, the strong veiling and limited S/N of the individual spectra do not allow us to constrain the magnetic field strengths. We therefore choose to perform the fit by assuming null magnetic fields. We indeed observe that the derived effective temperatures are not significantly affected by this choice. Appendix~\ref{appendix:plot_teff} provides an example of the fits obtained with this procedure.

Each spectrum was visually inspected, and only the fits obtained from the highest quality data were retained. When this method did not yield a reliable solution, due to excessive line broadening, low S/N, or strong veiling, we adopted another method. We have computed synthetic spectra using the \texttt{MARCS} atmospheric models and \texttt{SYNTH3} radiative transfer code. We used the Vienna Atomic Line Database\footnote{https://vald.astro.uu.se/} (VALD) to compute spectra over the SPIRou wavelength range. We have adopted microturbulent and macroturbulent velocities of 2\,\kms, solar abundances, and a surface gravity of $\log g=3.5$\,(cgs). We have computed a grid of models of effective temperatures between 3500 and 6000\,K with a step of 250\,K, and of \vsini\ values from 5 to 80 \kms\ with a step of 5\,\kms. For each observation we have selected one or two spectral regions in the H band, which gather spectral lines of enough S/N, and sensitive to the effective temperature. We used the independently determined veiling and radial velocity values, and then computed a $\chi^2$ value for each model. The effective temperature was selected based on the $\chi^2$ minimisation criterion. This method has the advantage to not fit too many parameters that cannot be constrained with the quality of our data, but to constrain the temperature with uncertainties around 250\,K, which is enough for the computation of the least-squares deconvolution (LSD) profiles (see Sect.~\ref{sec:lsd}). A variation of $\pm$500 K of the mask does not significantly change the LSD profile shapes, the \bl\ values, and the corresponding uncertainties. We emphasise that the quality of our data does not allow us to constrain the microturbulent velocity, and we therefore choose to fix it to usual values for young cool stars, around 1 or 2 \kms\ \citep{Doppmann2005}. As the broadening of the spectral lines is dominated by \vsini\ (see Table \ref{tab:vals}), the choice of the microturbulent velocity value has a negligible effect on the temperature fit. Similarly, the macroturbulent velocity cannot be constrained with our data. It affects only the wings of the spectral lines, which are usually lost in the noise. Our choices of 0 to 2 \kms\ for the macroturbulent velocity have a negligible impact on the effective temperature determination.

The veiling in the H- and K- bands \footnote{veiling measurements in the Y and J bands were not possible due to noise because of dust absorption} ($r_{\rm H}$ and $r_{\rm K}$) has been measured using our spectra by comparing photospheric spectral lines with those of a cool star template of the same effective temperature, which is evolved enough so that circumstellar material is absent and cannot contaminate the spectrum any further. The method is described in \citet{Sousa2023}, and the measured veiling values are summarised in Table \ref{tab:vals}.

Determining the effective temperature of VV CrA SW is challenging due to the weakness of its photospheric lines. According to \citet{Prato2003}, the SW component does not differ significantly from the NE component. However, the colour-colour diagram (see Fig. 3 in \citealt{Prato2003}) suggests a slightly lower temperature. VV CrA SW is therefore expected to be in the range of 3500 and 4000~K. We adopt an intermediate value of 3750~K for the effective temperature. This choice has only a minor impact on the LSD profiles and remains consistent with the presence of photospheric features observed in the LSD profiles (see Appendix \ref{appendix:lsd}), indicating that the corresponding temperature is appropriate.

Our analyses are strongly impacted by the large broadening and strong veiling for several stars. To mitigate the impact on our temperature estimates, we inflated the error bars to reflect the dispersion of the results obtained over several nights, when possible. The final temperatures were rounded to the nearest 250~K, matching the temperature grid of the line masks used for the LSD analysis.
The adopted effective temperatures range from 3250~K to 5000~K with a median of $\sim$4000~K, and are summarised in Table \ref{tab:teff}.

\begin{table}[t!]
\centering
\caption{Effective temperatures of our sample. 
}
\begin{tabular}{l|cr|c}
\hline
\hline
\label{tab:teff}
Source ID & \te & \multicolumn{1}{c|}{$\sigma$} & Method \\
  & (K) & \multicolumn{1}{c|}{(K)} & \\
\hline
V806 Tau & 4000 & 150 & ZeeTurbo \\
MHO 3 & 3750 & 100 & ZeeTurbo  \\ 
MHO 2 & 3250 & 50 & ZeeTurbo  \\ 
IRAS 04292+2422 W & 4750 & 50 & ZeeTurbo  \\
IRAS 04113+2758 N & 3500 & 100 & ZeeTurbo \\
IRAS 04369+2539 & 5000 & 250 & SYNTH3  \\
HL Tau & 4000 & 250 & SYNTH3  \\
IRAS 16191-1936 & 4000 & 200 & ZeeTurbo  \\
L1689 SNO2 & 3500 & 100 & ZeeTurbo  \\ 
IRAS 03247+3001 & 4000 & 250 & SYNTH3 \\ 
V512 Per & 5000 & 250 & SYNTH3 \\ 
VV CrA SW & 3750 & -- & -- \\ 
VV CrA NE & 4000 & 250 & SYNTH3 \\ 
SVS 20 S & 4000 & 250 & SYNTH3  \\
\hline
\end{tabular}
\tablefoot{Effective temperatures (\te) and their uncertainties ($\sigma$) determined using either \texttt{ZeeTurbo} or \texttt{SYNTH3}.}
\end{table}

\subsection{Least-squares deconvolved profiles}
\label{sec:lsd}

The LSD is a multi-line technique for computing a pseudo-average line profile as a function of velocity, from a spectrum \citep{Donati1997}. By combining the signal contained in all the spectral lines simultaneously, this method allows us to significantly increase the S/N of our data. The circular polarisation induced by the Zeeman effect at the surface of protostars is expected to be of the order of a few $10^{-4}$ times the continuum intensity, for a typical average line-of-sight magnetic strength of a few hundred gauss, in a moderate rotator (\vsini$\sim$20\,\kms). This signal cannot be detected in individual lines of spectra, even with S/N of 100 to 200. The LSD method allows  an increase of a factor of a few tens in the S/N (depending on the spectral type, and on the level of circumstellar contamination in the spectra), enabling the detection of stellar magnetic fields.

The fundamental principle of the LSD method is based on the hypothesis that the observed spectrum is the result of the convolution of an intrinsic line profile, inherent to the star, with a mask that indicates the wavelength, depth, and magnetic sensitivity of all the lines expected to form in the stellar spectrum. The method works by essentially reversing this process, i.e. deconvolving the observed spectrum by the line mask.

The line masks are generated using VALD \citep{Vald1995}, which selects the lines that can be formed in the photosphere of a star for a given effective temperature, surface gravity, microturbulence, and chemical composition. We used the effective temperatures tabulated in Table \ref{tab:teff}.
We assumed a \logg\ of 3.5, expected for Class~I sources \citep{Flores2024}, solar abundances, and a microturbulent velocity of 2\,\kms\ that is typical in stars with convective envelopes \citep{Doppmann2005}. 

In the following, we used the tools provided by the \texttt{SpecpolFlow} package \footnote{https://github.com/folsomcp/specpolFlow} developed and maintained by \citet{Folsom2025}. 
Before proceeding with LSD, the interactive mask cleaning tool \texttt{cleanMaskUI} was used to clean and tweak the line mask of each source. 
In the cleaning process, we removed the lines from the original mask that do not comply with the LSD assumption that all spectral lines have the same shape (e.g. emission lines or heavily damped lines like H lines). Additionally, we removed regions strongly contaminated by emission lines, the CO region (CO rovibrational lines from 2.29 $\mu$m and above), and the Y and J bands because they were too noisy in our sources. The tweaking process consists of adjusting the relative strengths of the lines of the mask to those observed, within one order. Careful cleaning and tweaking of the mask for each star significantly improve the LSD profile. Finally, to compute the profiles, we used the \texttt{LSDpy} package of \texttt{SpecpolFlow}.

Each LSD profile is computed within $\pm$ 200\ \kms\ around the central velocity of the individual lines, with a velocity pixel size of 2.4\,\kms\ to match the mean spacing of the spectral pixels, or 4.8\,\kms\ for the fastest rotators (\vsini\,$>\,$35\,\kms). For all sources, we used a magnetic Landé factor of 1.2, an average depth of 0.2 and a mean wavelength of 1660 nm, as normalisation factors.

The S/N of the LSD profiles is significantly higher than the S/N of the individual spectra, with a gain ranging from 20 to 75 (see Table \ref{tab:log}). The LSD profiles for the $I$, $V$, and $N$ spectra are shown in Fig. \ref{fig:lsd}.

\subsection{Magnetic field detection}

\begin{figure}[t!]
\centering
\includegraphics[width=1.0\linewidth]{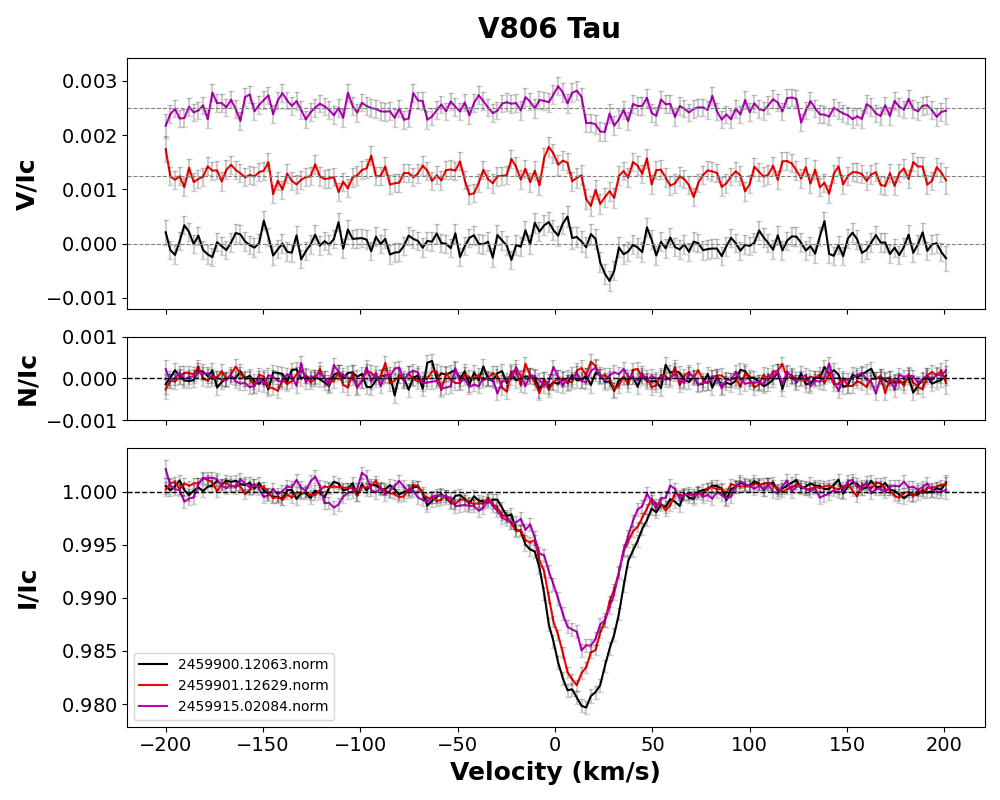}
\caption{\label{fig:exlsd}LSD profiles of V806 Tau. Different colours are used for the different observations. The Y-axes are I/Ic, N/Ic, and V/Ic, where Ic is the intensity of the stellar continuum. The $I$ profile represents the total intensity. Stokes $V$ shows a clear Zeeman signature, which corresponds to an actual magnetic detection because the signal is significantly non-null compared to the surrounding noise and the $N$ profile remains flat in the velocity domain.}
\end{figure}

A Zeeman magnetic signature manifests as a coherent non-zero signal spanning the full width of the photospheric line profile in the circular polarisation profile ($V$ profile) with no counterpart in the null profile ($N$ profile) as shown in Fig. \ref{fig:exlsd}. This ensures that the detected signal results from an astrophysical magnetic field, ruling out other phenomena that might influence the observations. 

We can observe in Fig. \ref{fig:lsd} that the $N$ profiles are consistent with zero, confirming that the polarimeter and data reduction perform as expected. If the $V$ profile shows a significant non-zero signal, while the $N$ profile remains null, we are confident that this signature is of stellar origin.

To assess the significance of the non-null Zeeman signature in the Stokes $V$ profile, we can derive the false alarm probability (FAP). It quantifies the likelihood that the observed signal is due to random fluctuations rather than a real physical phenomenon. To ensure robust results, the FAP should be as low as possible. Following \citet{Donati1997}, the criteria for interpreting the FAP and evaluating magnetic detections are: 

\begin{itemize}
    \item {Definite Detection}: the FAP must be below \(10^{-5}\), indicating a very low probability that the signal is due to noise, which means the magnetic field is detected.
    \item {Marginal Detection}: the FAP should be between \(10^{-5}\) and \(10^{-3}\), suggesting that the detection is probable but requires further verification or more observations to confirm it.
    \item {No Detection}: the FAP must be greater than \(10^{-3}\), indicating no magnetic field is detected. 
\end{itemize}

We computed the FAP as in \citet{Donati1997} using \texttt{SpecpolFlow}, by computing the reduced $\chi^2$ of the profile compared to a null model, and applied the $\chi^2$ statistics to convert them to detection probabilities ($P_{\rm det}$). The FAP is then computed as $1-P_{\rm det}$. This value depends on the uncertainties of the LSD $V$ profile. The cleaning and tweaking of the mask are therefore important. The FAP values are calculated within the photospheric LSD profiles. We therefore used a different velocity range for each star (see Table \ref{tab:limit}). Those ranges are defined as the regions where the LSD profile intensity is below the continuum in Stokes $I$. We have also computed the FAP in the $N$ profiles using the same velocity range, as a sanity check. If a detection is obtained in $N$ therefore the $V$ profile is not reliable, and no conclusion can be made. However, it is not the case in our sample. The FAP values are listed in Table \ref{tab:vals}

We considered a magnetic field to be definitively detected in a star if we have at least one definite detection in one $V$ observation, with no detection in the associated $N$ observation. According to this detection criterion, our sample contains five magnetic stars (V806 Tau, MHO 3, MHO 2, IRAS 16191-1936, L1689 SNO2), while nine stars do not show magnetic detection. Magnetic field appears to be detected only in FS protostars, this point is discussed in Sect.~\ref{sec:bias}.

\subsection{Mean longitudinal magnetic field (\( B_\ell \))}

The mean longitudinal magnetic field component (\( B_\ell \)) represents the projection of the magnetic field along the observer's line of sight, integrated over the observed stellar surface. The mean longitudinal magnetic field \( B_\ell \) is determined, in gauss (G), using the equation of \citet{Wade2000}:
\begin{equation}
B_{\ell} = -2.14 \times 10^{11} \frac{\int \frac{V(v)}{I_c}vdv}{\lambda z c \int \left[1 - \frac{I(v)}{I_c}\right] dv}.
\label{eq:bl}
\end{equation}
Here $\lambda$ is the central wavelength of the spectral line in nm, $z$ is the Landé factor, $c$ is the speed of light in vacuum, $v$ is the velocity, $V$ and $I$ are the Stokes parameters, and $I_{\rm c}$ is the intensity of the continuum. We apply this equation to LSD profiles. We have therefore used the normalisation values from the LSD computations: 1660 nm for $\lambda$ and 1.2 for $z$. We used the same velocity range as used for the FAP determination (Table \ref{tab:limit}). The LSD profiles are not shifted to the stellar rest frame, resulting in asymmetric velocity windows. The uncertainties on \bl\ are computed by propagating the errors of the $I$ and $V$ profiles in Eq. \ref{eq:bl}. The values of \bl\ are listed in Table \ref{tab:vals}. They range from $\sim$80\,G to $\sim$200\,G in absolute value for the five magnetic protostars.

\section{Discussion}

\label{sec:discussion}

\subsection{Study of observational bias}
\label{sec:bias}

In our final sample, six protostars (including V347 Aur) exhibit clear detections of large-scale magnetic fields, through Zeeman signatures in their Stokes $V$ profiles, confirmed with the FAP analysis. However, no magnetic field was detected in nine targets. This could indicate that these sources host significantly weaker magnetic fields, but it is first necessary to verify that this result is not driven by any observational or methodological bias. To better understand why some stars in our sample did not show detectable magnetic fields, we therefore investigated potential biases by comparing stars with definite magnetic detection (DD) to those with no detection (ND). We analysed how magnetic detectability correlates with various stellar and observational parameters, including the effective temperature, projected rotational velocity, veiling, signal-to-noise ratios, and the uncertainty on the longitudinal magnetic field measurements.

For each parameter, we constructed two cumulative distribution functions (CDFs), one for the DD sample, and one for the ND sample. We then performed a non-parametric Kolmogorov–Smirnov (KS) test to quantify the differences between the two distributions. The KS statistic \( s \) represents the maximum distance between the two CDFs, providing a measure of their discrepancy, while the p-value quantifies the statistical significance of this difference. A p-value of \( p > 0.05 \) indicates no significant difference, while \( p \leq 0.05 \) suggests a significant difference between the distributions. 

For some parameters such as \te, \vsini, and H mag, the tests are performed per source, while for the other parameters (e.g. veiling, $\sigma$(\bl), S/N) they are performed per observation. The source sample contains the 14 protostars analysed in Sect.~\ref{sec:analyse}, to which we added V347 Aur, analysed in \citet{Donati2024}, for which we used the observation of November 22nd 2022, representative of the 80 observations obtained for that source. The observation sample contains all observations listed in Table~\ref{tab:log} plus the V347 Aur observation. The veiling is not available for VV~CrA~SW because of the weak photospheric lines. The results of the tests are summarised in Table~\ref{tab:kstest}.

\begin{figure}[t!]
\centering
\includegraphics[width=1.0\linewidth]{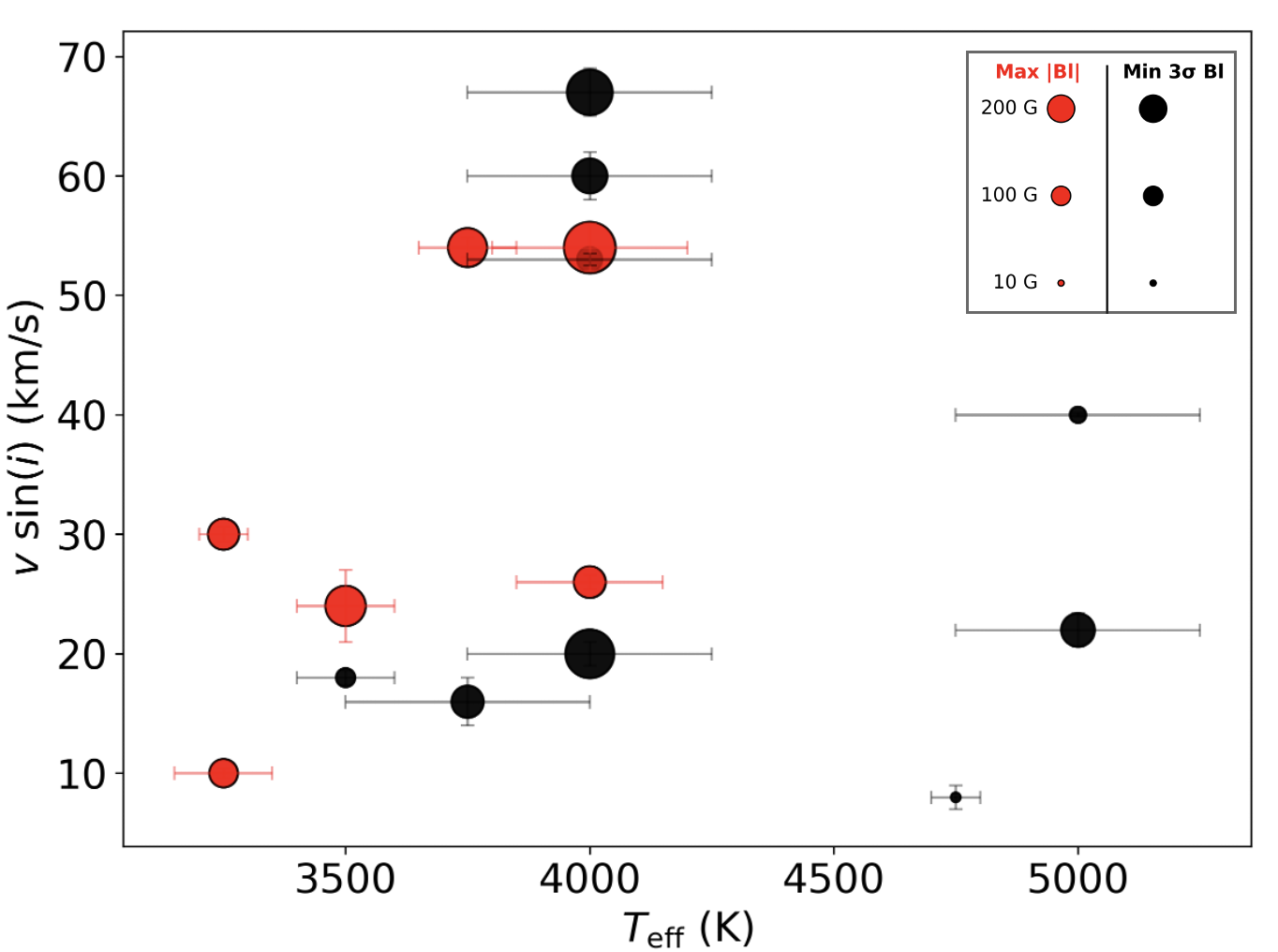}
\caption{\label{tempvsini}Distribution of projected rotational velocity for each star in the sample as a function of effective temperature. The magnetic stars are represented in red and the non-detected magnetic stars in black. For the magnetic stars, the sizes of the symbols are proportional to the maximum absolute values of the longitudinal magnetic field. For the non-detected magnetic stars, it is proportional to the minimum of the 3$\sigma$ uncertainty on \bl.
}
\end{figure}

On the birthline, at effective temperatures below $\sim$\,4750\,K (Fig.~1 in \citealt{Villebrun2019}), stars are expected to be fully convective at solar metallicity. For a given mass, as the temperature (hence age) increases, the convective envelope shrinks, possibly resulting in weaker fields, as observed in Class~II stars \citep{Gregory2012}. Furthermore, the higher the temperature, the lower the number of spectral lines contained in the mask. Thus, we could, expect a lower incidence of magnetic field detections in hotter stars. Nonetheless, we find that the effective temperature does not appear to significantly influence the occurrence of magnetic detections in our sample. In Fig.~\ref{tempvsini} we indeed observe that the detected magnetic stars are spread over our temperature range. However, this lack of correlation may not be significant, given the small sample size and the fact that our temperature range is relatively narrow.

\begin{table}[t!]
\caption{Results of the KS tests between the ND and DD samples.}
\label{tab:kstest}
\centering
\begin{tabular}{l | r r }
\hline
\hline
Tested  & \multicolumn{2}{c}{} \\
parameter & \multicolumn{1}{c}{$s$} & \multicolumn{1}{c}{$p$} \\
\hline
T$_{\rm eff}$ & 0.44 & 0.41 \\
\vsini & 0.28 & 0.90  \\ 
$r_{\rm H}$ & $\boldsymbol{0.70}$ & $\boldsymbol{0.0003}$ \\
$r_{\rm K}$ & $\boldsymbol{0.64}$ & $\boldsymbol{0.002}$ \\
$\sigma$(\bl) & $\boldsymbol{0.70}$ & $\boldsymbol{0.0004}$ \\  
S/N$_{\rm s}$ & 0.39 & 0.13 \\
S/N$_{\rm LSD}$ & 0.28 & 0.47 \\
H mag & 0.22 &  0.99 \\
\hline
\end{tabular}
\tablefoot{From  top to bottom the tested parameters are the effective temperature, projected rotational velocity, veiling in the H- and K- bands, uncertainty on \bl, the S/N in the spectrum and in the LSD profiles, and the magnitude in H. The $s$ and $p$ values of the KS tests are given between ND and DD. Bold fonts highlight the tests that significantly reject the null hypothesis at a 95\% confidence level (i.e. $p<0.05$).\
}
\end{table}

Our sample exhibits relatively simple Zeeman signatures, suggesting that the magnetic geometry is dominated by low-order large-scale magnetic field. The rotational broadening of the spectral lines could also affect magnetic detectability. For a similar magnetic field, the Stokes $V$ signature in a star with a low \vsini\ will have a larger amplitude than in a star with a large \vsini, as the same Zeeman signature will be spread over a smaller velocity range in the first case compared to the second one (presuming that the large-scale field is simple enough and regions of opposite polarities do not cancel out their respective signatures). If the S/N is the same in both cases, then the signal in the second case will be more diluted in the noise, and the FAP will be higher. When \vsini\ is larger, the magnetic detection probability is therefore lower.

Whatever the sample considered, we find that the null hypothesis cannot be rejected. Therefore, no significant difference is found between the DD and ND \vsini\ distributions. Remarkably, we are even able to detect magnetic fields in stars with \vsini\ as high as $\sim$55\,\kms. At the same time, large-scale magnetic fields are not detected in many protostars with low \vsini\ (Fig.~\ref{tempvsini}). Because, for almost all stars in our sample, the \vsini\ was unknown or poorly constrained at the time of the observations, we could not take this parameter into account in the computation of our exposure times. 
Our observational strategy  therefore appears  adequate as it allowed  a potential \vsini\ observational bias to be avoided.

\begin{figure}[t!]
\centering
\includegraphics[width=1.0\linewidth]{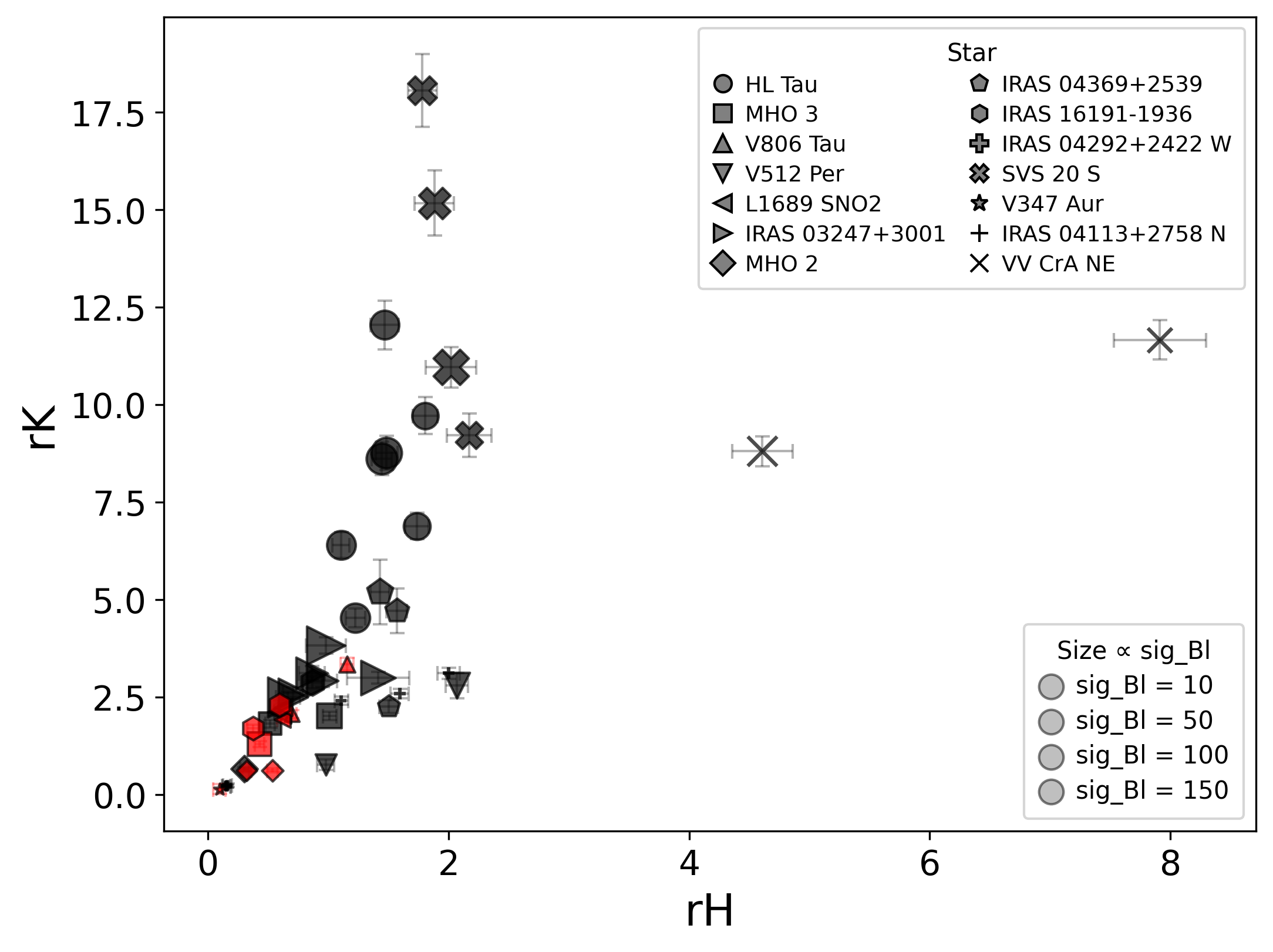}
\caption{\label{veiling}Veiling in the K band as a function of the veiling in the H band for all observations of the sample. The size of the symbol is proportional to the uncertainty of the longitudinal magnetic field and each symbol corresponds to a different target. Magnetic stars are represented in red and stars without magnetic detections in black.}
\end{figure}

Another parameter that may influence magnetic detections is the veiling. Veiling refers to the dilution of photospheric absorption lines due to excess continuum emission, typically interpreted as a signature of active accretion in young stars \citep{Connelley2010}. If a star exhibits strong veiling, this can affect the detectability of magnetic fields by reducing the line contrast with respect to the intensity continuum, and therefore increasing the relative noise in Stokes $I$ and $V$ profiles (see Appendix \ref{appendix:veiling}). The KS tests reject the null hypothesis at a confidence level higher than 95\% when comparing the veiling distributions of the DD and ND samples, in both H and K bands, and in both statistical samples. In Fig.~\ref{veiling}, we indeed find that the higher the veiling, the lower the detection rate. 

This indicates that the detectability of magnetic fields in our sample is influenced by the veiling. For some objects we have more observations than for others. We therefore verified that the veiling–detectability trend remains when these targets (HL~Tau and IRAS~03247+3001) are excluded. If the veiling is not known before setting up the exposure time of the observation, and because the depths of the photospheric lines in a veiled spectrum are shallower than in an unveiled one, the veiled photospheric line profiles will be noisier than the unveiled ones, which lowers the contrast of the Zeeman signatures and may hamper Stokes $V$ detections (see Appendix \ref{appendix:veiling}). The fact that we do not detect magnetic fields in the most veiled observations could therefore be an observational bias. Moreover, veiling in Class~I is higher than in Class~II, which could also explain the lower detectability in our sample of Class~I and FS protostars.

Figure \ref{veiling} illustrates that magnetic detections are found only for the lowest values of $r_{\rm H}$ and $r_{\rm K}$. The targets with the highest veiling are those for which the analysis of the photospheric lines is most challenging. A significant discrepancy between the adopted and the true effective temperature could also affect the derived veiling values. However, this would modify the veiling by at most a factor of two, which does not alter our conclusions, as the associated uncertainties remain smaller than the overall veiling amplitude observed across our sample. We also remark that magnetic detections are only observed in FS protostars, while none of the Class~I protostars is detected. Class I sources are usually more veiled than FS sources, which could explain this result.

Finally, we may also wonder whether the quality of our data affects magnetic detection. We find no significant differences between the S/N distributions of the DD and ND samples, considering the S/N in either the spectra (S/N$_{\rm s}$) or the LSD profiles (S/N$_{\rm LSD}$).

Our magnetic detection rate is not affected by the S/N of our data but is affected by veiling. This means that we should observe a correlation between the magnetic field detections and the uncertainties of the longitudinal field measurements (see Appendix \ref{appendix:veiling}). We find that the distributions of $\sigma$(\bl) for the DD and ND samples are significantly different (Table \ref{tab:kstest}). Furthermore, Fig.~\ref{veiling} shows that the uncertainties on \bl\ are systematically lower for the least veiled spectra. The veiling phenomenon could therefore be the main source of difficulty for detecting magnetic fields in Class I and FS sources.

\subsection{Upper limits on the dipole magnetic field}

Stars in which we did not detect a magnetic field may still host one at their surface, but it may be too weak or too complex to be detected with our current data quality. In this section we aim to estimate the maximal magnetic field strength a star can have that would still remain hidden in the noise of our data. In the case of a complex magnetic field, this is not easy because such fields can exhibit a large variety of topologies, and a large amplitude of strength, which could remain hidden within a given noise level. On the other hand, magnetic fields observed in fully convective stars (i.e. in CTTSs or M dwarfs) are mainly dipolar \citep{Donati2011, Gregory2012, Morin2008, Morin2010}. 
Furthermore, our sample mostly contains simple Zeeman signatures, most likely produced by low-order magnetic fields. We  therefore only consider a dipole to compute the upper limit.

To compute the upper limits on the magnetic field strength at the pole for each of these stars, we assume an oblique rotator \citep{Stibbs1950}, i.e. a star hosting a dipole of strength \Bpol\ at the pole, inclined by a magnetic obliquity $\beta$ with respect to the rotation axis $\Omega$, which is itself inclined by an angle $i$ with respect to the line of sight. For a given star with a given \vsini, and a given depth $d_I$ of the LSD $I$ profile, we can predict the shape of the Stokes $V$ profile at a given rotation phase, compare its amplitude to the noise of our observations, and use the FAP method to determine whether it would be detectable or not. 

The weak field approximation is assumed to compute the local Stokes $V$ profiles \citep{Landi2004}. We can then run Monte-Carlo simulations assuming random $i$, $\beta$, and rotation phase for one value of \Bpol, and count the number of detections. We repeat this process for several values of \Bpol, and trace the detection probability curve (e.g. Fig.~\ref{fig:upperlimit}). To be consistent with the literature, we chose the upper limit \Bpol\ at a detection probability of 90\%. The full description of the methodology is detailed in \citet{Alecian2016}, and recalled in Appendix~\ref{appendix:upperlimits}. 

We first computed the upper limits for the non-detected stars using marginal and definite FAP criteria. The resulting values are summarised in Table~\ref{tab:limit}. These upper limits are relatively high, ranging from approximately 500 G to more than 5 kG, suggesting that relatively strong dipoles might indeed be present in these stars, but are below the sensitivity of our observations. 

\begin{table}[t!]
\caption{Velocity range and upper limits on the dipolar magnetic field derived using the marginal (MD) and definite (DD) detection criteria for each star.}
\label{tab:limit}
\centering
\begin{tabular}{lrrr}
\hline
\hline
\multirow{2}{*}{Source ID} & {Velocity range} & \multicolumn{2}{c}{max(\Bpol) (G)} \\
 & (\kms) & (MD) & {(DD)} \\
\hline
\textbf{V806 Tau} &  $-30 / 40$ & 1900 & 3700\\
\textbf{MHO 3} &  $-50 / 60$  & 4900 & >5000\\ 
\textbf{MHO 2} &   $-16 / 40$ & 3000 & 4300\\ 
IRAS 04292+2422 W &   $15 / 40$ & 480 & 650\\
IRAS 04113+2758 N &   $-6 / 30$ & 1900 & 2300\\
IRAS 04369+2539 &    $-30 / 60$ & >5000 & >5000\\
HL Tau &   $-40 / 65$ & 4900 & >5000\\
\textbf{IRAS 16191-1936} &   $-30 / 40$ & >5000 & >5000\\
\textbf{L1689 SNO2} &  $-45 / 25$ & 2500 & 4100\\ 
IRAS 03247+3001 &  $-60 / 60$  & >5000 & >5000\\ 
V512 Per &  $-10 / 30$ & 3300 & 4100\\ 
VV CrA SW &  $-20 / 20$ & >5000 & >5000 \\ 
VV CrA NE &  $-25 / 20$ & 4700 & >5000\\ 
SVS 20 S &  $-70 / 60$ & >5000 & >5000 \\
\hline
\end{tabular}
\tablefoot{The detected magnetic protostars are indicated in bold font.}
\end{table}

To assess the reliability of these upper limit values, we also computed them for stars with detected magnetic fields. We find values between approximately 2\,kG and more than 5\,kG. We observe that the highest upper limits can be found independently in detected and non-detected magnetic protostars. The projected rotational velocity affects the upper limit on the dipole field strength. A higher \vsini\ dilutes the Stokes $V$ signature, implying that stronger magnetic fields are required for a detection in stars with large \vsini, compared to stars with low \vsini. We find that all sources with a limit above 5 kilogauss have a \vsini\ larger than 35 \kms. The fact that such large limits are found in DD and ND protostars means that large \vsini\ is not the main parameter that hampers the detection of magnetic field. This is in line with the similar \vsini\ distributions that we find between the DD and ND samples (Sect.~\ref{sec:bias}).

It is interesting to note that both ND and DD samples have similar upper limits. This suggests that if our non-detected sample hosts dipolar magnetic fields of comparable strength to those of the detected stars, they would likely have been detected. The non-detected stars might then host weaker or more complex magnetic fields, if any.

\subsection{Impact of our magnetic measurements on the origin and evolution of stellar magnetic fields}

We conducted a spectropolarimetric analysis of a sample of Class~I and FS protostars, aiming to investigate the presence of large-scale magnetic fields. All young, low-mass observed protostars are expected to be fully convective or to have a large convective envelope \citep{Palla1998,Baraffe2017}, we expect all of them to host detectable magnetic fields, generated by convective dynamos, as in low-mass main-sequence stars. However, magnetic fields were detected in only 6 out of the 15 stars of our final sample, corresponding to 40\,\% of the sample. This relatively low detection rate raises several questions regarding the magnetic origin of these young stars, their internal processes, and their magnetospheric environments. 
In particular, the absence or apparent non-detection of large-scale magnetic fields in a substantial fraction of Class~I and FS objects challenges the assumption that such fields are ubiquitous at these evolutionary stages. 

A central question that emerges is whether large-scale magnetic fields, commonly observed in more evolved systems such as CTTSs, are truly universal across all largely convective stellar objects. Assuming that our protostars lie on the birthline, we expect them to be fully convective up to an effective temperature of 4750~K, and have a convective envelope of mass larger than 40\,\% of the stellar mass up to 5250~K \citep{Villebrun2019}. Under this assumption, $\sim$80\,\% of our sample is fully convective, while the remainder has a radiative core plus a large convective envelope. We can also estimate the range of mass probed by our sample using the effective temperatures derived in Sect.~\ref{sub:temp}. We find a range of 0.2 to 2.5\,\msun. This means that their direct descendants are the CTTSs of low- and intermediate-mass \citep{Herbig1962,Villebrun2019}. The magnetic properties of these objects have been intensely characterised thanks to dense spectropolarimetric monitoring, and ZDI analysis \citep[e.g.][]{Hussain2009,Donati2013,Pouilly2020,Nowacki2023,Zaire2024}. Magnetic fields are detected in fully convective stars, and almost all of them host a strong ($>$\,1\,kG) magnetic field, dominated by a dipolar component, and sometimes a multipolar magnetic field. When published, the maximum absolute \bl\ values are larger than 400\,G and can reach 900\,G. Our snapshot observations do not allow us to perform ZDI to retrieve the topology of the magnetic field we have detected in our sample. However, we do not detect strong magnetic fields in all of them: the maximum absolute \bl\ value is 200\,G. 
A single measurement of \bl\ is strongly dependent on the rotation phase. Our phase sampling is less complete than that of the ZDI studies, and therefore we may have missed the rotation phases when \bl\ is maximum. However, out of the 16 observations we obtained for the 5 magnetic protostars, none are higher than 200 \,G, which are at most similar, but mainly weaker than in fully convective CTTSs \citep{DonatiTW, DonatiCI, Nowacki2023}.

On one hand, Class I and FS protostars have a larger radius than CTTSs as they are less evolved. 
Class I and FS protostars are younger than 1 Myr, but probably older than 0.5 Myr \citep{Fiorellino2023}. 
CTTSs are 1 to 3 Myr old \citep[e.g.][]{Zaire2024}. According to stellar evolutionary models \citep[e.g.][]{Baraffe2015}, from 0.5 Myr to 3 Myr, the radius decreases by a factor up to 2. If we assume a constant magnetic flux as the protostars evolve towards the pre-main sequence, then the surface magnetic field can increase by a factor up to 4, which could explain why the \bl\ in our sample are in general weaker than at later stages. 
On the other hand, it is not clear whether magnetic flux conservation can be invoked when magnetic fields are of dynamo origin. The major structural difference between Class~I and fully convective Class~II stars is the surface gravity, hence the gravity gradient within the convection zone. 
Magneto-hydro-dynamical models of convective stars of various stratifications are required to understand the role of stratification in the efficiency of dynamo processes. Within the PROMETHEE project, we are developing models for different stratification levels that will help us to interpret this result (Guseva et al., in prep.).

Finally, it is interesting to mention that in fully convective main-sequence M dwarfs, a phenomenon known as dynamo bistability has been observed \citep{Morin2010}. In this scenario, stars with identical global properties (mass, rotation rate, convection regime) can exhibit either strong predominantly dipolar magnetic fields or much weaker, complex, and multipolar fields. It is plausible that a similar mechanism is at play in our sample, and could explain part of the non-detections.

\subsection{Comparison with the total magnetic fields}

A study of 32 Class~I and FS protostars was conducted by \citet{Flores2024}, focussing on measuring the average strength of the total magnetic field. They used K-band observations with the high-resolution near-infrared echelle spectrograph iSHELL on the IRTF (Hawaii), achieving a spectral resolution of 50\,000. Their targets share similar star-forming regions, magnitudes, and spectral indices with ours. They used the radiative transfer code Moog Stokes \citep{Deen2013} to derive stellar parameters (effective temperature, surface gravity, mean magnetic field strength <$B$>, and projected rotational velocity \vsini ), as well as stellar masses and ages, based on magnetic stellar evolutionary models from \citet{Feiden2016}.

\begin{figure}[h]
\centering
\includegraphics[width=1.0\linewidth]{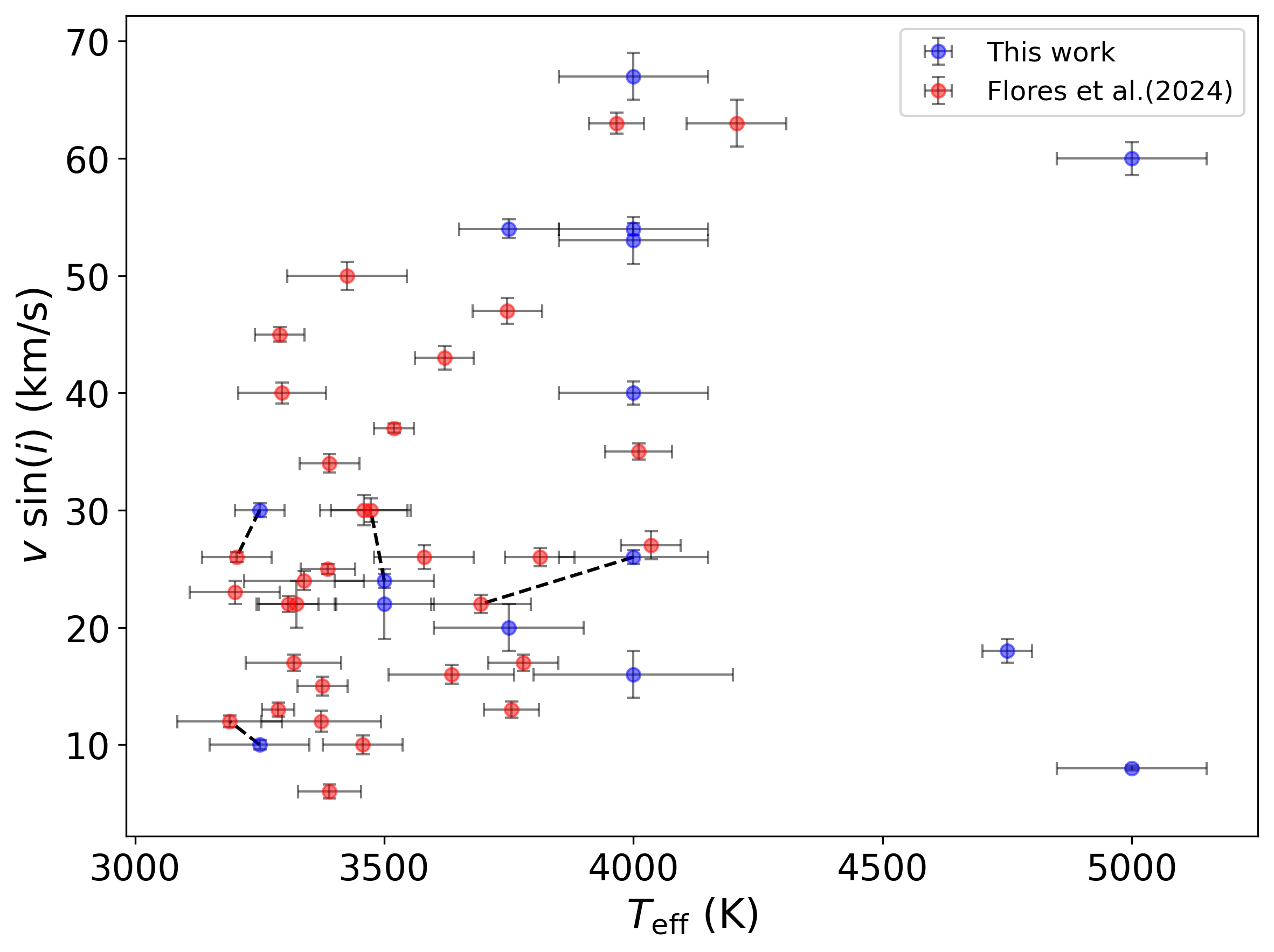}
\caption{\label{fig:florescomparison}Projected rotational velocities \vsini as a function of effective temperatures measured in this work (blue) and in the work of \citet[red]{Flores2024}. The  points connected by the dashed line indicate the common stars in the two samples.}
\end{figure}

As their data are of similar quality to ours, and they used a model similar to ours to fit the spectra, it is first interesting to compare their results with ours. Both studies cover similar effective temperatures, and \vsini\ ranges (see Fig.~\ref{fig:florescomparison}). \citet{Flores2024} reports a mean effective temperature of 3450 K and an average \vsini\ of 26\,\kms, while our sample shows a higher average effective temperature of 4000 K and an average \vsini\ of 36 \kms\ . The distributions of the two samples are therefore slightly different. In proportion, we have more stars with \vsini\ larger than 40\,\kms, and with effective temperatures higher than 4000\,K.
In addition, we find that we have four targets in common with their sample: MHO 2, V806 Tau, L1689 SNO2, and V347 Aur, which enable a direct comparison. We find similar projected rotational velocities, \vsini, although our derived effective temperatures are slightly higher, but still consistent within the uncertainties.

Out of their 32 sources, \citet{Flores2024} was able to measure non-null <$B$> in 27 of them, and report upper limits for the 5 remaining sources between 0.8 and 1.2\,kG. Among the 27 sources with a measured magnetic field, they report an average magnetic field strength of approximately 2 kG, consistent with the values typically found in more evolved Class~II sources \citep{JohnsKrull2007}. These measurements, based on Zeeman broadening, are sensitive to the total magnetic field, including both large and small-scale components. In contrast, our spectropolarimetric method is sensitive only to large-scale magnetic fields. These two approaches probe different components. While Zeeman broadening captures the cumulative surface magnetic field regardless of orientation, spectropolarimetry detects only the longitudinal component of the field. This means that regions of opposite polarity on the stellar surface can cancel each other out. If the stellar magnetic field is only organised on small scales, or is a combination of large- and small-scales fields,  the spectropolarimetric method leads to an underestimation of the total magnetic field \citep{See2019, Lehmann2019}. 

As already mentioned, our snapshot sample cannot be used to determine the topology and strength of the large scale magnetic field, and therefore estimate its contribution to the total magnetic field measured through Zeeman broadening. However, if we assume that our magnetic sample contains only dipolar fields, then we can use the statistical relationship between the cumulative longitudinal magnetic field (<$B_{\ell}$>$_{\rm rms}$) of a magnetic sample, and its average total magnetic strength <$B$>$_{\rm av}$. \citet{Oleg2024} determined a statistical relationship between the two parameters by running many simulations of dipoles at the surface of an A0 main-sequence star for random magnetic obliquities and inclinations, and of a linear limb-darkening parameter of 0.5. The average value of limb-darkening parameter of our sample is 0.4 \citep{Claret2011}. When correcting for our different limb-darkening value, the relationship can be written as
\begin{equation}
\label{eq:bav}
\langle B \rangle_{\mathrm{av}} = 4.04^{+1.1}_{-0.5} \, \langle B_{\ell}\rangle_{\rm rms},
\end{equation}

where $\langle B_{\ell}\rangle_{\rm rms}$ is the root-mean-square of our measurements of \bl\ in our magnetic sample.

Based on Eq. \ref{eq:bav}, the mean surface field modulus for our magnetic sample would be approximately 450 ± 200 G, assuming it hosts only dipolar fields. 

This value is well below the average value obtained from Zeeman broadening by \citet{Flores2024}, as in most low-mass stars, i.e. consistent with dynamo fields. This suggests that a significant portion of the magnetic energy in these protostars is stored in small-scale magnetic structures, likely concentrated in active regions or spots, rather than in a globally organised dipole.

Complementary observations, including the monitoring of some targets, will allow us to better measure and characterise the full magnetic topology through ZDI. This method will provide a more detailed view of the large-scale field and a more complete picture of the magnetic environments in Class~I and FS protostars.

When they are strong enough, large-scale fields are responsible for funnelling accretion flows, generating hot spots, and regulating angular momentum transport through magnetospheric interaction \citep{Bouvier2014, Romanova2015}. Meanwhile, small-scale magnetic activity contributes to coronal heating, X-ray emission, and potentially to disk ionisation and chemistry, impacting planet formation conditions from the earliest evolutionary stage of solar-type stars  \citep{Grosso2020}. 
Therefore, characterising both the small and large-scale magnetic fields is essential to build a coherent picture of the magnetic environment in protostars. Such an analysis will be presented for a few stars in a following paper.

\section{Conclusions}
\label{sec:ccl}

We have carried out a detailed analysis of 14 Class~I and FS protostars using SPIRou, aiming at characterising their stellar and large-scale magnetic properties. When including V347~Aur, the final statistical sample comprises 15 objects. When possible, we derived effective temperatures, and using LSD profiles, we measured the detection probability of a Zeeman signature in the Stokes $V$ profiles, the longitudinal magnetic fields, the projected rotational velocities, and the radial velocities. Our results can be summarised as follows:

   \begin{enumerate}
      \item We detected large-scale magnetic fields in 6 out of 15 protostars, with absolute longitudinal field strengths ranging from approximately 80\,G to 200\,G. In contrast, nine objects show no clear magnetic signature.
      \item The effective temperatures for our sample of Class~I and FS sources range from 3250 K to 5000 K, with a median value of approximately 4000~K.
      \item We find no significant correlation between magnetic field detection and stellar parameters, such as effective temperature or projected rotational velocity.
      \item We find a correlation between magnetic field detection and veiling, suggesting that veiling influences the detectability of magnetic fields by diluting the signal in the Stokes $V$ profile.
      \item We derived upper limits on the dipolar field strength, with a median value below 3 kG, indicating that strong magnetic fields may still be present, but are below our detection threshold when no magnetic field is detected.
      \item The longitudinal magnetic field values of our magnetic sample do not seem to reach the higher levels observed in their evolutionary descendants, i.e. the fully convective CTTSs. 
      \item Our statistical estimate of the average total field strength of our magnetic sample is significantly lower than those obtained via Zeeman broadening studies (e.g. \citealt{Flores2024}). This may suggest that a significant portion of the magnetic energy in these protostars is stored in small-scale magnetic structures, likely concentrated in active regions or spots, rather than in a globally organised dipole. 
   \end{enumerate}

Our analysis highlights the challenges of detecting large-scale magnetic fields in Class I and FS protostars, especially when the fields are weak, complex, or below our detection thresholds. This naturally raises the question of how early in the stellar formation process magnetic fields become detectable and dynamically important. 

In CTTSs, the magnetic star–disk interaction is well documented \citep{Bouvier2007, Hartmann2016}. The strong dipolar component of the stellar magnetic field truncates the inner disk at a few stellar radii, and funnels accreting material along magnetic field lines, forming accretion columns. In a following paper, we will investigate whether there is any correlation between magnetic field detections and the equivalent widths of Br$\gamma$ and Pa$\beta$ emission lines, which are commonly used as accretion tracers in embedded protostars \citep{Antoniucci2008, Fiorellino2021}. 
We cannot rule out that magnetospheric accretion processes are at play in the all sources that we observed. If magnetospheric accretion occurs in Class~I and FS objects, strong kilogauss large-scale magnetic fields are likely required, given the higher mass accretion rates of such objects compared to Class~II \citep{Fiorellino2023}.

The comparison with Zeeman broadening measurements, which are sensitive to the total surface magnetic field, suggests that a significant fraction of the magnetic energy in these young stars may reside in small-scale field structures, undetectable through circular polarisation due to spatial cancellation, or that the magnetic topologies may vary widely, possibly due to dynamo bistability, as seen in fully convective M dwarfs \citep{Morin2010}. This underlines the importance of combining different observational diagnostics to obtain a comprehensive view of stellar magnetism.

Importantly, recent studies have pushed the observational frontier to even earlier stages. Although Class 0 protostars are even more deeply embedded and challenging to observe, recent studies have begun to detect photospheric absorption lines in these objects \citep{Greene2018, LeGouellec2025}. These detections open promising prospects for future spectropolarimetric studies of Class 0 sources, offering the potential to provide critical constraints on the initial conditions of dynamo generation and accretion physics, and  trace magnetic field evolution from the very first stages of stellar formation. Class~0 objects have  accretion diagnostics that are similar to those in the more evolved Class I and II protostars \citep{Laos2021,LeGouellec2024}, perhaps suggesting that strong and organised stellar magnetic fields are present.

This study demonstrates the capabilities of SPIRou to probe the large-scale magnetic fields of embedded protostars, making it a powerful tool to characterise magnetism in the early stages of star formation. The next step is to perform ZDI on our magnetic sample.
This will  unveil the topology of the magnetic field in Class~I and FS objects, which has not been done to date. These topologies will be a key element in understanding the evolution of magnetic fields in protostars and of the star--disk interactions.

\begin{acknowledgements}
We want to thank the referee for helpful comments which
significantly improved the clarity of the manuscript.
This work has received funding from the French Agence Nationale de la Recherche (ANR) through the project PROMETHEE (ANR-22-CE31-0020). This research has made use of the SIMBAD database, the VizieR catalogue access tool, and the Aladin sky atlas, CDS, Strasbourg Astronomical Observatory, France. OK acknowledges support by the Swedish Research Council (grant agreement no. 2023-03667) and the Swedish National Space Agency. AK acknowledges support by the NKFIH NKKP grant ADVANCED 149943 and the NKFIH excellence grant TKP2021-NKTA-64. Project no.149943 has been implemented with the support provided by the Ministry of Culture and Innovation of Hungary from the National Research, Development and Innovation Fund, financed under the NKKP ADVANCED funding scheme. CPF acknowledges funding from the European Union's Horizon Europe research and innovation programme under grant agreement No. 101079231 (EXOHOST) and from UK Research and Innovation (UKRI) under the UK government's Horizon Europe funding guarantee (grant number 10051045). H.N. acknowledges funding from the European Research Council (ERC) under the European Union’s Horizon 2020 research and innovation programme (Grant agreement No. 101019653). V.J.M.L.G. acknowledges support by the Spanish program Unidad de Excelencia María de Maeztu CEX2020-001058-M, financed by MCIN/AEI/10.13039/501100011033, and by the MaX-CSIC Excellence Award MaX4-SOMMA-ICE.
V.J.M.L.G. acknowledges support by the European Research Council (ERC) under the European Union’s Horizon 2020 research and innovation program (grant agreement No. 101098309 - PEBBLES).
\end{acknowledgements}

\bibliographystyle{aa}
\bibliography{biblio}

\appendix

\section{The dipole strength upper limit method}
\label{appendix:upperlimits}

We implemented a method to derive the projected rotational velocity, line depth, and radial velocity from observed stellar absorption line profiles by fitting synthetic profiles computed from a physical stellar model. The model computes spectral line profiles based on the three key parameters. These parameters are iteratively updated in a configuration file before each model run. The synthetic profiles are compared to observed data by computing the total \(\chi^2\) statistic, accounting for measurement uncertainties. A Nelder–Mead simplex algorithm is employed to minimise \(\chi^2\) by adjusting the parameters simultaneously, searching best-fit values for line depth, \vsini\, and \vrad. Confidence intervals for each parameter are derived through one-dimensional \(\chi^2\) scans. For each parameter, values around the optimum are scanned while re-optimising the others, and uncertainties correspond to \(\Delta \chi^2 = 1\), representing 68 per cent confidence intervals.

The Nelder-Mead algorithm is particularly well suited for minimising functions that are difficult to differentiate, such as a \(\chi^2\) function. 
To better constrain the initial parameters, we estimated them with a Gaussian fit, which provides a reliable estimate of the most probable initial values. To estimate the uncertainties associated with the optimised parameters, we performed a detailed \(\chi^2\) analysis. Starting from the optimal parameter values, each parameter was varied individually by 20\% around its best-fit value while keeping the other parameters fixed. For each parameter, the \(\chi^2\) function was evaluated at 20 evenly spaced points within this range. The minimum total \(\chi^2\) value is computed, the upper and lower bounds of the parameter were found by locating the points where \(\chi^2 = \chi^2_{\text{min}} + 1\). These bounds define the asymmetric uncertainties (\(\sigma_-\) and \(\sigma_+\)) for each parameter, reflecting the sensitivity of the \(\chi^2\) function to changes in the parameter values. To further validate the reliability of the uncertainties, we compared the results with other methods, such as Monte Carlo simulations and Probability Density Function analysis. Additionally to the one-dimensional analysis, we performed a two-dimensional \(\chi^2\) analysis to examine correlations between pairs of parameters. In this case, the criterion for the 68\% confidence region corresponds to \(\Delta \chi^2 = 2.4\), as there are two free parameters. 
These complementary approaches confirmed the robustness of the uncertainty estimates, ensuring that the reported confidence intervals are both accurate and reliable.

\begin{figure}[t!]
\centering
\includegraphics[width=1.0\linewidth]{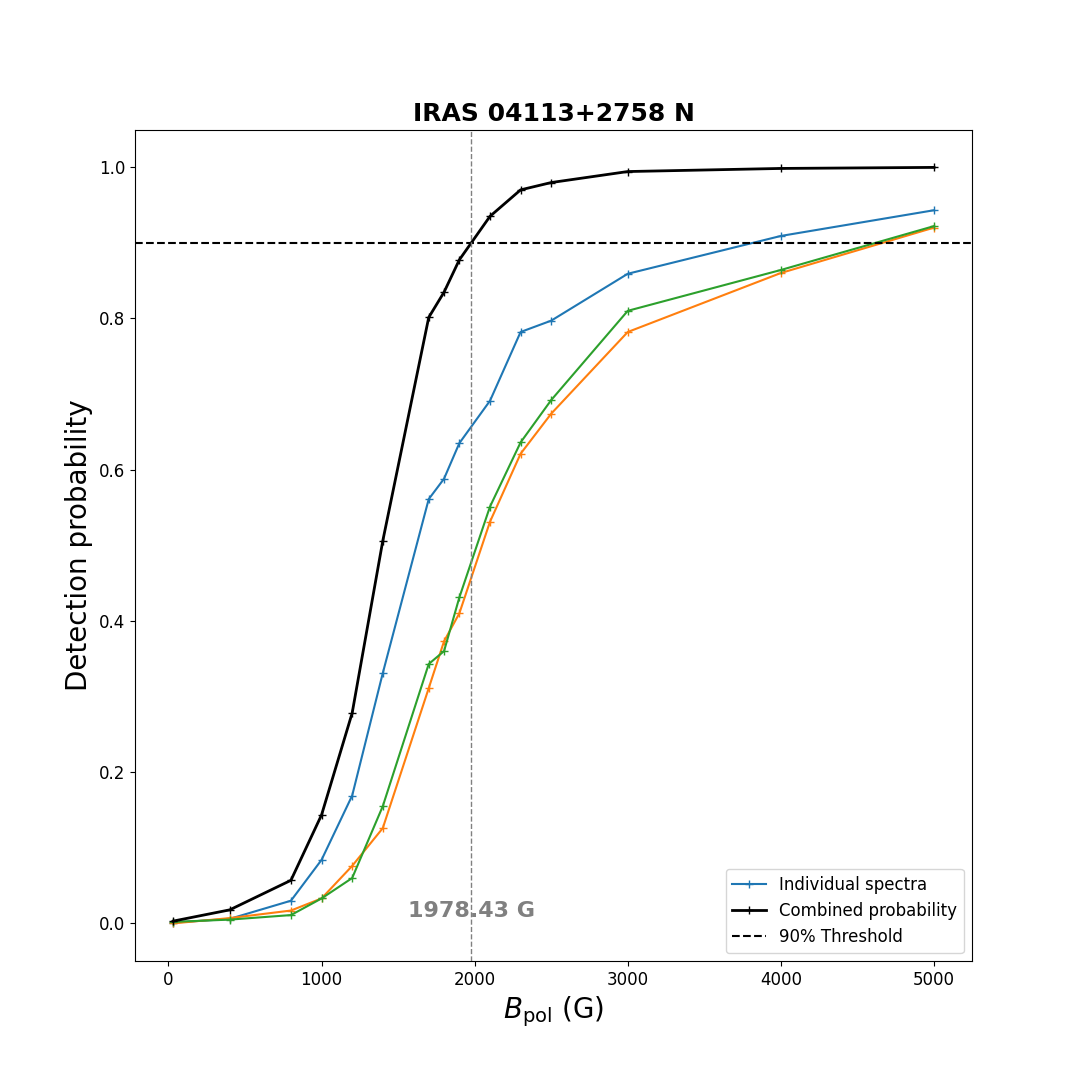} 
\hfill
\includegraphics[width=1.0\linewidth]{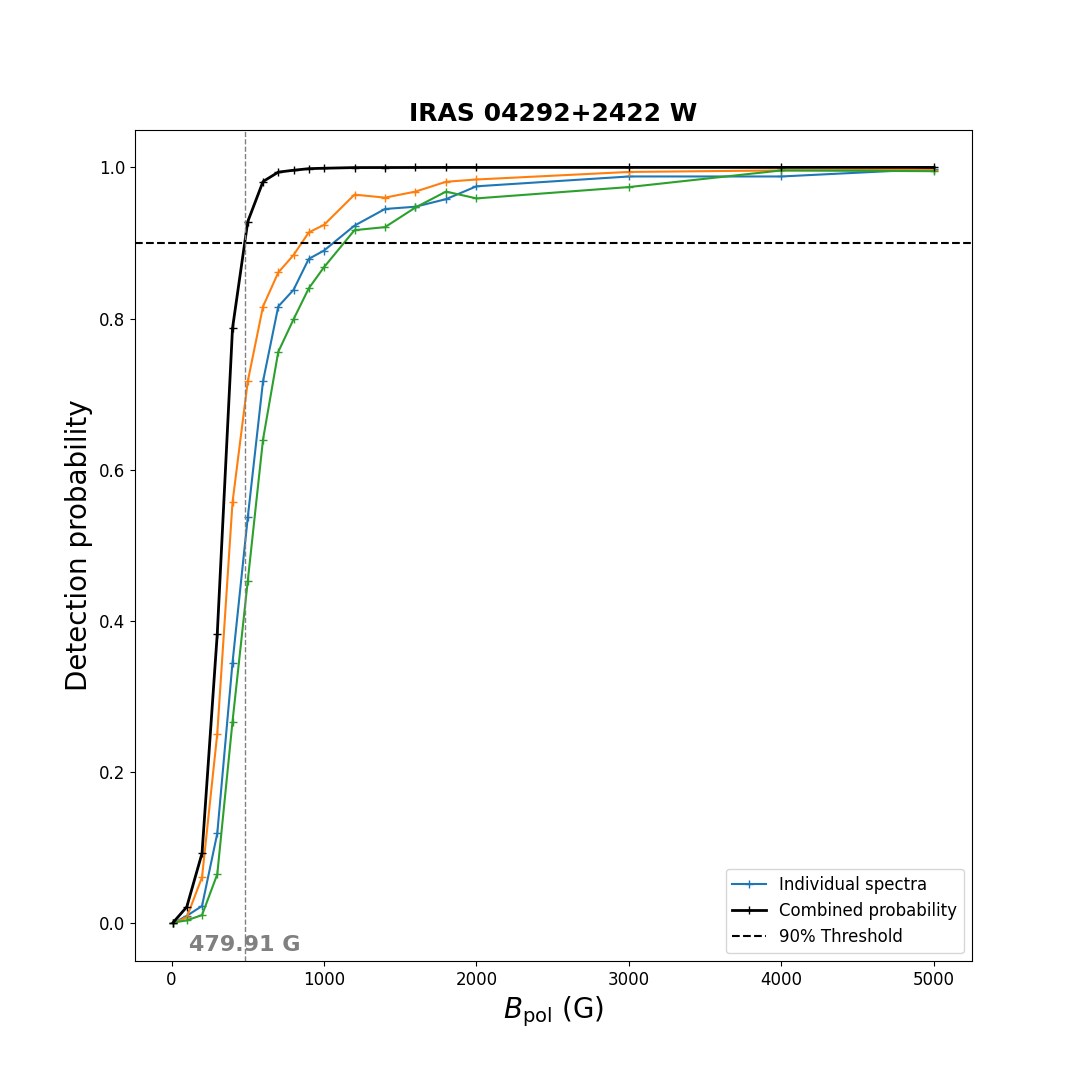} 
\caption{\label{fig:upperlimit}Example of dipolar magnetic detection probability curve as a function of the dipole strength. The coloured curves represent the probability for each individual observation, while the black curve represents the combined probability for all observations. The 90 per cent threshold is used to determine the field strength upper limit.}
\end{figure}

The detection limit of the magnetic field, denoted \( B_{\text{lim}} \), represents the value below which the magnetic field signal can remain below the noise level \citep{Alecian2009,Alecian2016}. To estimate the maximum strength, we use the oblique rotator model \citep{Stibbs1950}. We assume a dipolar configuration, which is the simplest magnetic model (see Fig.~9 in \citealt{Alecian2016}), and fit the corresponding local profiles to our LSD $I$ observations. The three free parameters in the fitting procedure are the $I$ line depth, the \vsini, and the radial velocity. We fitted each observation independently, which results in slightly different values of these parameters for different observations of the same star. However, the measurements remain consistent within the error bars. The variations observed in \vsini\ can be explained by line profile variability in the Stokes $I$ profiles (see Fig~\ref{fig:lsd}).

Next, a Monte-Carlo method is used to simulate a grid of 1000 synthetic $V$ profiles for each observation, where random values are assigned to the model parameters. Each model uses a random inclination angle $i$, obliquity angle $\beta$, and rotational phase $\Omega$, with various values of the polar magnetic fields \( B_{\text{pol}} \) using the same mean Landé factor (1.2) and wavelength (1660 nm). The local line profiles are computed using an instrumental broadening corresponding to a resolving power of 70\,000 and a macroturbulent velocity of 5 \kms. The shapes of the wings of the LSD $I$ profiles are not compatible with a null macroturbulent velocity. Furthermore, if we let it be free in the fit, we obtain values that are unphysical. We therefore made a compromise and chose to fix it to a value that reproduces well the wings of the LSD profiles, but is not too high. We should caution that the computation of the LSD profiles themselves may produce artificially broadened wings, for example because of line blending, or may produce depressed continuum.
It  can therefore be higher than the value needed to fit the individual spectral lines for the temperature determination in Sect. \ref{sub:temp}. This value should therefore be interpreted with caution. The surface integration of the local profiles is done using a linear limb-darkening coefficient. These values have been extracted from the \citet{Claret2011} database for the corresponding effective temperatures of our stars (Sect.~\ref{sub:temp}), assuming a surface gravity of 3.5, and solar abundances. Because our LSD masks are dominated by H-band lines, we choose the coefficients computed in this band, and we used the mean values of both methods (least-squares or flux conservation) used by \citet{Claret2011} to compute them.

Gaussian noise is added to each profile. This process is repeated for all the observations, with the 1000 models of $V$ profiles enabling the calculation of the detection probability for a range of magnetic field strengths. For each set of models, the probability of detecting a magnetic field in each simulated $V$ profile is computed. This allows us to determine whether a magnetic field is detected or not, using the FAP. We ensured that the FAP does not exceed a prescribed value, using FAP = \(10^{-3}\) for marginal detection \citep{Donati1997}. We chose the marginal criterion to be consistent with previous publications, and be able to compare with other samples. We also compute the upper limit using definite criterion. However, the choice of this criterion is arbitrary, as all values will be affected in a similar way. For each value of the field strength, we calculated the number of detections across several models for each observation and plotted the detection rates as a function of field strength. A detection rate of 90 per cent is required to statistically consider that the field would be detected in one observation. This enables us to determine the upper limit of the field strength, below which we are unable to detect it in our observations. Assuming the observations are all independent, we can compute the combined probability following the method outlined by \citet{Neiner2015}, for various polar field strengths, and apply the 90 per cent detection threshold to derive the upper limit. By analysing this curve, we can determine a value of \( B_{\text{lim}} \). This corresponds either to a profile containing only noise or to one in which the signature of a weak magnetic field is hidden within the spectral noise in Stokes V (see Fig~\ref{fig:upperlimit}). 

\section{Fitting example with ZeeTurbo}
\label{appendix:plot_teff}

The best fit for one observation of IRAS 04113+2758 N, showing all the lines used in the temperature fitting.

\begin{figure*}[t!]
\centering
\includegraphics[height=12cm]{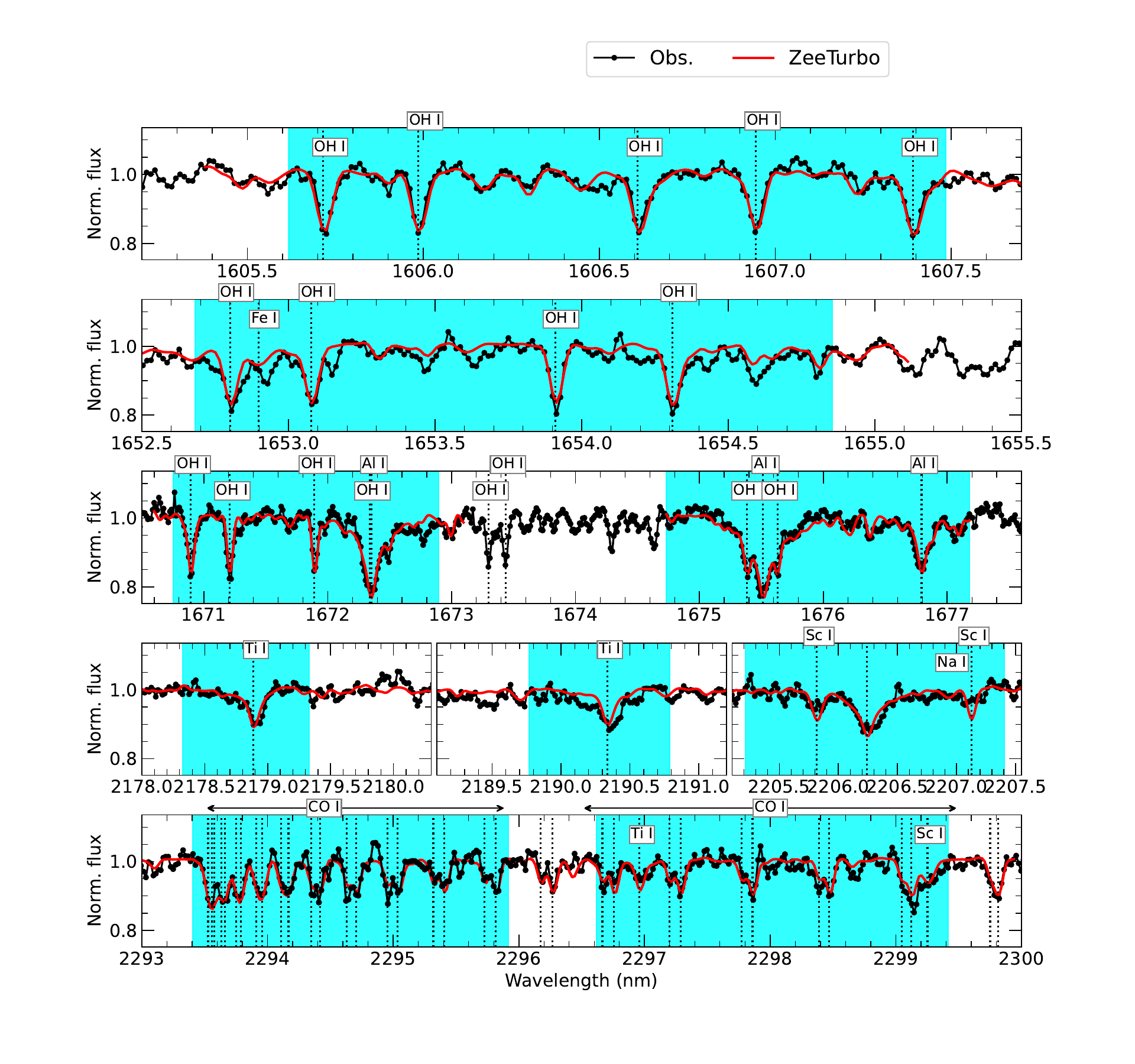}
\caption{\label{fig:fit}ZeeTurbo fitting process for the first observation of IRAS 04113+2758 N. The observed spectrum is shown in black, and the best-fit model in red. The blue shading indicates the regions used for parameter fitting, and the corresponding absorption lines are labelled.}
\end{figure*}

\section{The effect of veiling on the uncertainties of \bl}
\label{appendix:veiling}

As our sources can be strongly affected by veiling, i.e. a continuum excess that adds to the stellar continuum, we explore its effect on the uncertainties on the \bl\ measurement. We can re-write Eq. (\ref{eq:bl}) as
\begin{alignat*}{3}
\hspace{2em}&B_{\ell}=\alpha\frac{X}{Y}&\hspace{4em}&\text{where}&\hspace{4em}&X=\int \frac{V(v)}{I_c}vdv\\ 
\hspace{2em}&&\hspace{4em}&&\hspace{4em}&Y=\int \left[1 - \frac{I(v)}{I_c}\right] dv
\end{alignat*}
and $\alpha=-2.14\times10^{11}/(\lambda z  c) \approx -358$\,G\,km$^{-1}$\,s\ when using the normalisation values of $\lambda$ and $z$ of our LSD profiles. The numerical computation of $X$ and $Y$, and of their uncertainties $\sigma_X$ and $\sigma_Y$ can be written
\begin{alignat*}{2}
\hspace{1em}&X=\sum^N_{i=1}\left(\frac{V}{I_c}\right)_iv_i\Delta v_i&
\hspace{2em}&\sigma_X=\sqrt{\sum^N_{i=1}\left[\left(\frac{\sigma_V}{I_{\rm c}}\right)_iv_i\Delta v_i\right]^2}\\
\hspace{1em}&Y=\sum^N_{i=1}\left[1 - \left(\frac{I}{I_c}\right)_i \right]\Delta v_i&
\hspace{2em}&\sigma_Y=\sqrt{\sum^N_{i=1}\left[\left(\frac{\sigma_I}{I_{\rm c}}\right)_i\Delta v_i\right]^2}
\end{alignat*}
where $\sigma_I$ and $\sigma_V$ are the uncertainties on the unnormalised $I$ and $V$ spectra, and $N$ is the number of pixels within the photospheric profile. The uncertainty on \bl\ is therefore
\begin{alignat*}{1}
\hspace{1em}&\frac{\sigma_B}{B_{\ell}}=\sqrt{\left(\frac{\sigma_X}{X}\right)^2+\left(\frac{\sigma_Y}{Y}\right)^2}
\end{alignat*}

We consider a star that we observe twice. During the first observation the star is not affected by veiling, and emits a spectrum $I_1(v)=I_0(v)$ with a continuum $I_{\rm 1c}=I_{\star}$. During the second observation the star is contaminated with an additional excess continuum $I_{\rm ex}(v)=rI_{\star}(v)$, where $r$ is defined as the veiling parameter. We  assume that $r$ does not depend on the wavelength, and that the region emitting the veiling does not emit circular polarisation. During the second observations we therefore record the spectrum $I_2(v)=I_0(v)
+I_{\rm ex}(v)$, with a continuum $I_{\rm 2c}=I_{\star}+I_{\rm ex}$. Both polarised spectra are assumed identical, and an associated polarised spectrum $V_2(v)=V_1(v)=V(v)$. $X$ and $Y$ for both observations are therefore
\begin{alignat*}{4}
\hspace{1em}&X_1&=&\sum^N_{i=1}\left(\frac{V}{I_{\star}}\right)_iv_i\Delta v_i&
\hspace{2em}&X_2&=&\sum^N_{i=1}\left(\frac{V}{I_{\star}+I_{\rm ex}}\right)_iv_i\Delta v_i\\
\hspace{1em}&Y_1&=&\sum^N_{i=1}\left[1 - \left(\frac{I}{I_{\star}}\right)_i \right]\Delta v_i&
\hspace{2em}&Y_2&=&\sum^N_{i=1}\left[1 - \left(\frac{I+I_{\rm ex}}{I_{\star}+I_{\rm ex}}\right)_i \right]\Delta v_i\\
\hspace{1em}& &=&\sum^N_{i=1}\left(\frac{I_{\star}-I}{I_{\star}}\right)_i \Delta v_i& 
\hspace{2em}& &=&\sum^N_{i=1}\left(\frac{I_{\star}-I}{I_{\star}+I_{\rm ex}}\right)_i \Delta v_i
\end{alignat*}
As the continuum in the LSD profiles is not dependent on $v$, when dividing $X$ by $Y$ in both observations, the numerators cancel, and $X_1/Y_1=X_2/Y_2$, hence 
\begin{equation*}
\hspace{1em}B_{\ell1}=B_{\ell2}=B_{\ell}    .
\end{equation*}

To explore how veiling affects the computation of the uncertainties, we assume first that both observations were obtained in identical weather conditions, with the same exposure time. The noise is therefore identical in both observations: $\sigma_{I1}=\sigma_{I2}=\sigma_I$, and $\sigma_{V1}=\sigma_{V2}=\sigma_V$
\begin{alignat*}{4}
\hspace{1em}&\sigma_{X1}&=&\sqrt{\sum^N_{i=1}\left[\left(\frac{\sigma_V}{I_{\star}}\right)_iv_i\Delta v_i\right]^2}&
\hspace{2em}&\sigma_{X2}&=&\sqrt{\sum^N_{i=1}\left[\left(\frac{\sigma_V}{I_{\star}+I_{\rm ex}}\right)_iv_i\Delta v_i\right]^2}\\
\hspace{1em}&\sigma_{Y1}&=&\sqrt{\sum^N_{i=1}\left[\left(\frac{\sigma_I}{I_{\star}}\right)_i\Delta v_i\right]^2}&
\hspace{2em}&\sigma_{Y2}&=&\sqrt{\sum^N_{i=1}\left[\left(\frac{\sigma_I}{I_{\star}+I_{\rm ex}}\right)_i\Delta v_i\right]^2}
\end{alignat*}
When dividing $\sigma_X$ by $X$ and $\sigma_Y$ by $Y$, the denominators of $X$ and $Y$ cancel, and therefore
\begin{equation*}
\hspace{1em}\frac{\sigma_{B1}}{B_{\ell1}}=\frac{\sigma_{B2}}{B_{\ell2}}=\frac{\sigma_B}{B_{\ell}}.
\end{equation*}
We note that in that case, in the second observation the received signal is higher than in the first observation. Its S/N is therefore higher than that of the first one: $(S/N)_2=(S/N)_1 \sqrt{1+r}$. However, because we received the same flux from the star itself, it is normal that the uncertainty on \bl\ is not affected.

We now assume a second case where we observe two identical stars of the same magnitude. One is veiled, and the other one is not. However, before scheduling the observation, we do not know that the second star is veiled. We therefore observe them with the same exposure time, in the same weather conditions. Because the magnitudes are the same, the recorded fluxes for both stars are the same, and therefore $(S/N)_2=(S/N)_1$. The uncertainties on the flux from the stars, however, are not the same: 
\begin{equation*}
\hspace{1em}\sigma_{I2} = \sigma_{I1}\sqrt{1+r}, \qquad \sigma_{V2} = \sigma_{V1}\sqrt{1+r}.
\end{equation*}
Substituting these expressions into the definitions of $\sigma_X$ and $\sigma_Y$ give
\begin{alignat*}{4}
\hspace{0em}&\sigma_{X1}&=&\sqrt{\sum^N_{i=1}\left[\left(\frac{\sigma_{V1}}{I_{\star}}\right)_iv_i\Delta v_i\right]^2}&
\hspace{1em}&\sigma_{X2}&=&\sqrt{\sum^N_{i=1}\left[\frac{\sqrt{1+r}}{(1+r)}\left(\frac{\sigma_{V1}}{I_{\star}}\right)_iv_i\Delta v_i\right]^2}\\
\hspace{0em}& & & &
\hspace{1em}& &=&\frac{\sigma_{X1}}{\sqrt{1+r}}\\
\hspace{0em}&\sigma_{Y1}&=&\sqrt{\sum^N_{i=1}\left[\left(\frac{\sigma_{I1}}{I_{\star}}\right)_i\Delta v_i\right]^2}&
\hspace{1em}&\sigma_{Y2}&=&\sqrt{\sum^N_{i=1}\left[\frac{\sqrt{1+r}}{(1+r)}\left(\frac{\sigma_{I1}}{I_{\star}}\right)_i\Delta v_i\right]^2}\\
\hspace{0em}& & & &
\hspace{1em}& &=&\frac{\sigma_{Y1}}{\sqrt{1+r}}
\end{alignat*}
Hence, for the veiled observation we obtain 
\begin{equation*}
\hspace{1em}\frac{\sigma_{X2}}{X_2} = {\sqrt{1+r}}\frac{\sigma_{X1}}{X_1}, \qquad
    \frac{\sigma_{Y2}}{Y_2} = {\sqrt{1+r}}\frac{\sigma_{Y1}}{Y_1},
\end{equation*}
and 
\begin{equation*}
\hspace{1em}\frac{\sigma_{B2}}{B_{\ell2}}= {\sqrt{1+r}} \frac{\sigma_{B1}}{B_{\ell1}}.
\end{equation*}

Therefore, the uncertainty on \bl\ increases by a factor $\sqrt{1+r}$ compared to the unveiled case. 

In summary, when we know the unveiled magnitude of a star, we can set up the exposure time to get sensitive measurements of the magnetic field, and in that case, the uncertainty on \bl\ is unaffected by veiling if we don't change the exposure time, whatever is the flux magnitude during the observation. On the contrary, if the H- or K-band magnitudes of a veiled star are used to set up the exposure time, then the uncertainty on \bl\ is multiplied by a factor $\sqrt{1+r}$ compared to the unveiled star. Anyhow, the measured values of \bl\ remain unchanged in both cases.

\section{Log of observations}
\label{appendix:SPIROU_log}

Table. \ref{tab:log} summaries the log of observations obtained with SPIRou spectropolarimeter.

\begin{table*}[t!]
\caption{Log of the SPIRou observations.}
\small
\label{tab:log}
\centering
\begin{tabular}{@{}l c r@{\;\;}c@{\;} r c c r r r@{}}
\hline
\hline
Source ID & RA & \multicolumn{1}{c}{DEC} & SFR  & \multicolumn{1}{c}{H} & Date & BJD & \multicolumn{1}{c}{Exp.} & \multicolumn{1}{c}{S/N} & \multicolumn{1}{c}{S/N}\\
 &  (hh:mm:ss) & \multicolumn{1}{c}{(deg:mm:ss)} &  & \multicolumn{1}{c}{(mag)} &  & -2400000 & \multicolumn{1}{c}{Time (s)} &  \multicolumn{1}{c}{@1.67\,$\mu$m} & \multicolumn{1}{c}{LSD}\\
\hline
\object{V806 Tau*} & 04:32:15.42 & +24:28:59.6 & Taurus & 9.3 & 2022-11-16 & 59900.12063 & 1850 & 174 & 7257 \\
 & & & & & 2022-11-17 & 59901.12629 & 1850 & 180 & 6783 \\
 & & & & & 2022-12-01 & 59915.02084 & 1850 & 198 & 7575 \\
\hline
\object{MHO 3*} & 04:14:30.55 & +28:05:14.6 & Taurus & 9.2 & 2022-11-16 & 59900.09753 & 1738 & 106 & 4450 \\
 & & & & & 2022-12-01 & 59914.97152 & 1738 & 130 & 6230 \\
 & & & & & 2022-12-04 & 59918.01608 & 1738 & 158 & 7735 \\
 & & & & & 2023-01-01 & 59945.96152 & 1738 & 126 & 5986 \\
\hline
\object{MHO 2*} & 04:14:26.40 & +28:05:59.6 & Taurus & 9.4 & 2022-12-04 & 59918.04028 & 2050 & 182 & 7054 \\
 & & & & & 2023-01-02 & 59946.87428 & 2050 & 164 & 6948 \\
 & & & & & 2023-01-03 & 59947.85486 & 2050 & 170 & 6687 \\
\hline
\object{IRAS 04292+2422 W} & 04:32:13.27 & +24:29:10.8 & Taurus & 9.9 & 2023-01-02 & 59946.90951 & 3209 & 164 & 6299 \\
 & & & & & 2023-01-02 & 59947.89613 & 3209 & 162 & 6703 \\
 & & & & & 2023-01-03 & 59952.97509 & 3209 & 168 & 5635 \\
\hline
\object{IRAS 04113+2758 N*} & 04:14:26.30 & +28:06:03.3 & Taurus & 9.9 & 2023-01-04 & 59948.93997 & 3120 & 156 & 7372 \\
 & & & & & 2023-01-09 & 59953.85844 & 3120 & 148 & 8150 \\
 & & & & & 2023-01-10 & 59954.83884 & 3120 & 152 & 7622 \\
\hline
\object{IRAS 04369+2539*} & 04:39:55.75 & +25:45:01.9 & Taurus & 8.1 & 2022-11-19 & 59903.13177 & 602 & 150 & 5407 \\
& & & & & 2022-11-21 & 59905.05519 & 602 & 118 & 4956 \\
& & & & & 2022-12-01 & 59915.06549 & 602 & 158 & 5798 \\
\hline
\object{HL Tau} & 04:31:38.45 & +18:13:57.2 & Taurus & 9.2 & 2018-09-22 & 58384.11335 & 2407 & 126 & 6876 \\
& & & & & 2018-09-23 & 58385.09837 & 2407 & 126 & 5983 \\
& & & & & 2018-09-25 & 58387.08102 & 2407 & 108 & 5644 \\
& & & & & 2019-01-16 & 58499.80432 & 2407 & 124 & 7167 \\
& & & & & 2019-01-17 & 58500.79189 & 2407 & 132 & 6048 \\
& & & & & 2020-09-18 & 59111.08910 & 3610 & 170 & 8873 \\
& & & & & 2020-09-22 & 59115.05352 & 3610 & 196 & 10014 \\
\hline
\object{IRAS 16191-1936*} & 16:22:04.35 & -19:43:26.7 & Ophiuchus & 9.7 & 2024-03-18 & 60388.10393 & 3209 & 102 & 3509 \\
 & & & & & 2024-05-02 & 60433.02290 & 3209 & 106 & 3978 \\
 & & & & & 2024-06-26 & 60487.76569 & 3209 & 118 & 5145 \\
\hline
\object{L1689 SNO2*} & 16:31:52.12 & -24:56:15.7 & Ophiuchus & 9.8 & 2019-05-14 & 58618.10070 & 1404 & 92 & 4780 \\
& & & & & 2024-07-18 & 60509.89855 & 3499 & 168 & 7536 \\
& & & & & 2024-07-26 & 60517.85402 & 3499 & 158 & 6886 \\
\hline
\object{GY92 214} & 16:27:09.40 & -24:37:18.8 & Ophiuchus & 11.5 & 2019-05-14 & 58618.01656 & 1203 & 21 & 2432 \\
& & & & & 2019-05-14 & 58618.03393 & 1203 & 22 & 2765 \\
\hline
\object{GY92 378} &  16:27:51.80 & -24:31:45.5 & Ophiuchus & 12.5 & 2019-05-14 & 58618.07152 & 3611 & 54 & 2752 \\
\hline
\object{IRAS 03247+3001*} & 03:27:47.68 & +30:12:04.5 &  Perseus & 9.4 & 2023-01-02 & 59946.84774 & 2095 & 96 & 3149 \\
& & & & & 2023-01-03 & 59947.81412 & 2095 & 94 & 2836 \\
& & & & & 2023-01-04 & 59948.84853 & 2095 & 100 & 3199 \\
& & & & & 2023-01-09 & 59953.90247 & 4012 & 148 & 3892 \\
& & & & & 2023-01-10 & 59954.88269 & 4012 & 130 & 3570 \\
& & & & & 2023-01-11 & 59955.89845 & 4012 & 148 & 4934 \\
\hline
\object{IRAS 03413+3202*} & 03:44:32.00 & +32:11:43.9 & Perseus & 9.4 & 2022-12-03 & 59943.90751 & 2095 & 124 & 4257 \\
& & & & & 2022-12-30 & 59948.95959 & 2095 & 122 & 4324\\
\hline
\object{V512 Per} & 03:29:03.76 & +31:16:04.0 & Perseus & 9.7 & 2020-08-28 & 59090.03171 & 4814 & 98 & 1947 \\
& & & & & 2023-01-04 & 59948.87769 & 2541 & 62 & 3264 \\
\hline
\object{VV CrA SW} & 19:03:06.7 & -37:12:49.6 & Corona Australis & 8.4 & 2024-04-01 & 60402.15170 & 980 & 164 & 7771 \\
& & & & & 2024-04-24 & 60425.12513 & 980 & 132 & 6698 \\
& & & & & 2024-06-22 & 60484.02208 & 980 & 112 & 6099 \\
& & & & & 2024-06-23 & 60484.96903 & 980 & 108 & 5868 \\
\hline
\object{VV CrA NE*} & 19:03:06.6 & -37:12:49.7 & Corona Australis & 9.7 & 2024-04-25 & 60426.07900 & 3164 & 119 & 5730 \\
& & & & & 2024-06-23 & 60484.99885 & 3164 & 116 & 6713 \\
& & & & & 2024-06-24 & 60485.98446 & 3164 & 97 & 4196 \\
\hline
\object{SVS 20 S*} & 18:29:57.7 & +01:14:05.7 &  Serpens & 9.2 & 2024-04-20 & 60430.99864 & 2652 & 126 & 7825 \\
& & & & & 2024-06-26 & 60487.99729 & 3165 & 164 & 12068 \\
& & & & & 2024-06-27 & 60488.91144 & 3165 & 178 & 11276 \\
& & & & & 2024-06-28 & 60489.91152 & 3165 & 132 & 9392 \\
\hline
\end{tabular}
\tablefoot{The columns indicate (from  left to the right) the identification number (ID) of the source (when two components were resolved by the telescope, we indicate the pointed target using its cardinal direction with respect to its companion); barycentric right ascension (RA) and declination (DEC) in the ICRS (Ep=2016.0) reference system given by \citet{Gaia2022}; star-forming region (SFR); H-band magnitude; date of observation; barycentric Julian date (BJD); total exposure time for one $V$ sequence; peak S/N in spectral pixels of  order 34 (1.67 $\mu$m); and S/N in the LSD profiles. Sources with an * symbol are flat-spectrum targets.}
\end{table*}

\section{LSD Profiles and parameter values}
\label{appendix:lsd}

The LSD profiles of each star are shown in Fig.~\ref{fig:lsd}. 
Table. \ref{tab:vals} summaries for each observation, the corresponding values for the depth $I$, \vsini\ , \vrad\ , \bl\ , and if the magnetic field is detected or not.

\begin{figure*}[h]
\centering
\includegraphics[width=1.0\linewidth]{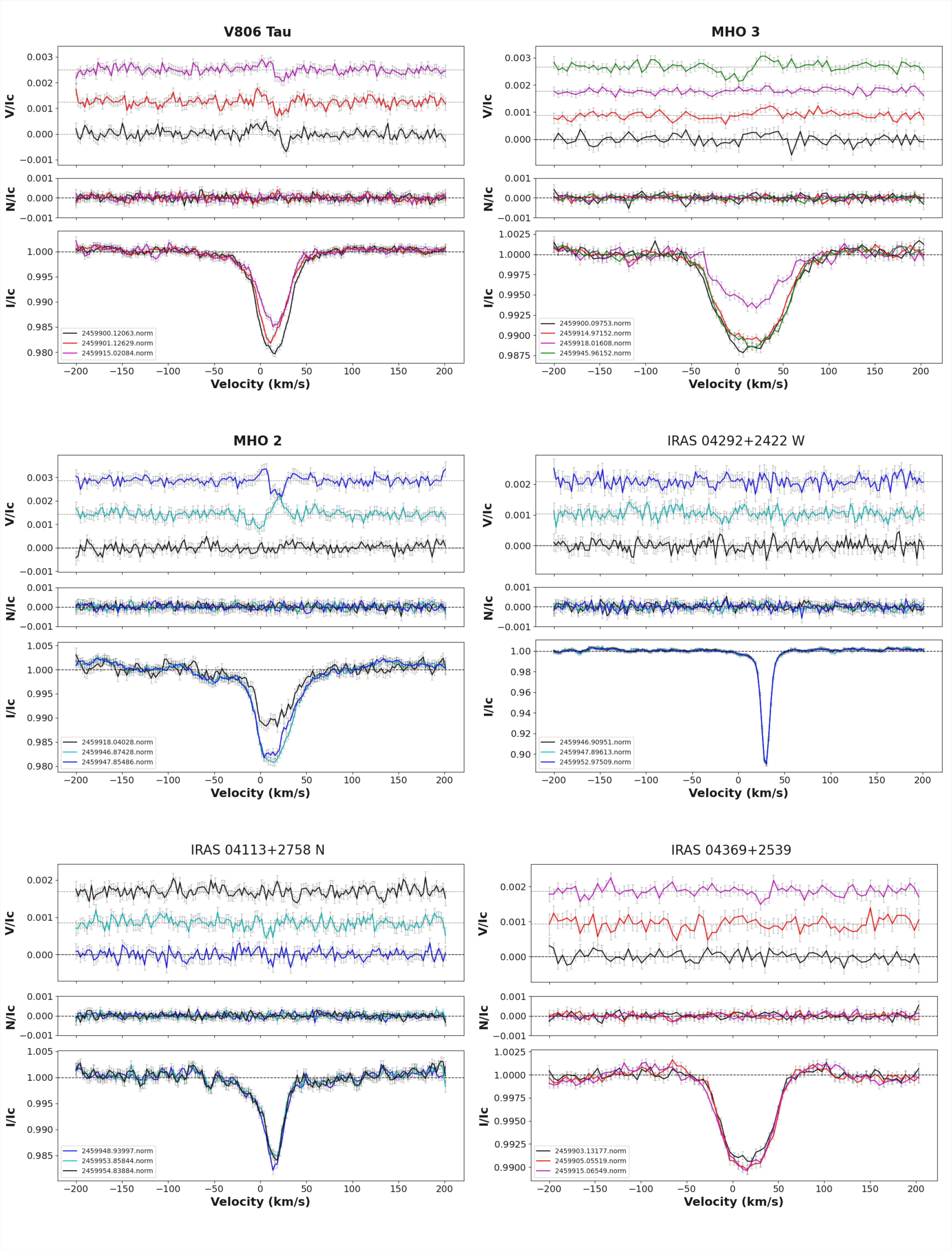}
\end{figure*}

\begin{figure*}[h]
\centering
\includegraphics[width=1.0\linewidth]{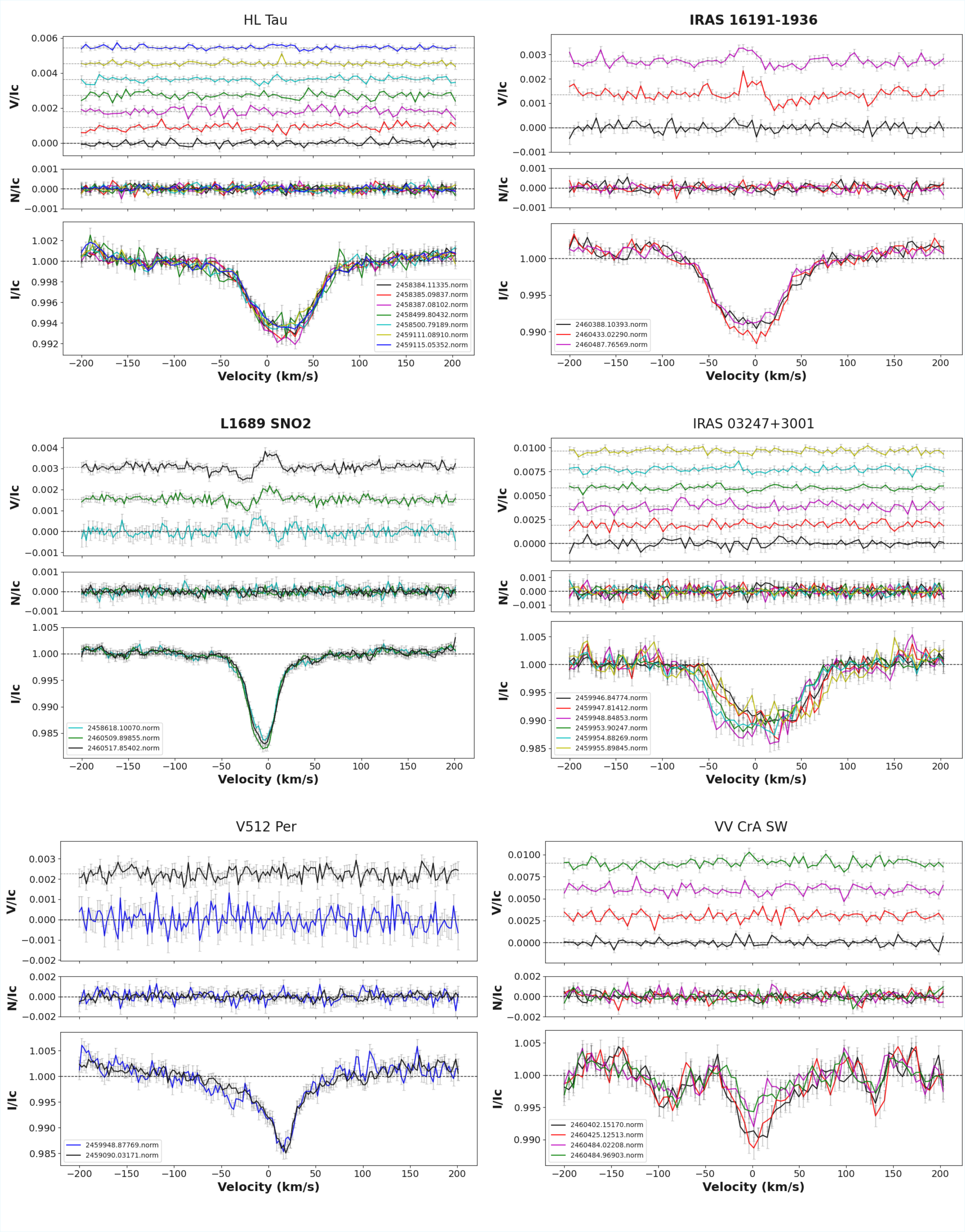}
\end{figure*}

\begin{figure*}[h]
\includegraphics[width=1.0\linewidth]{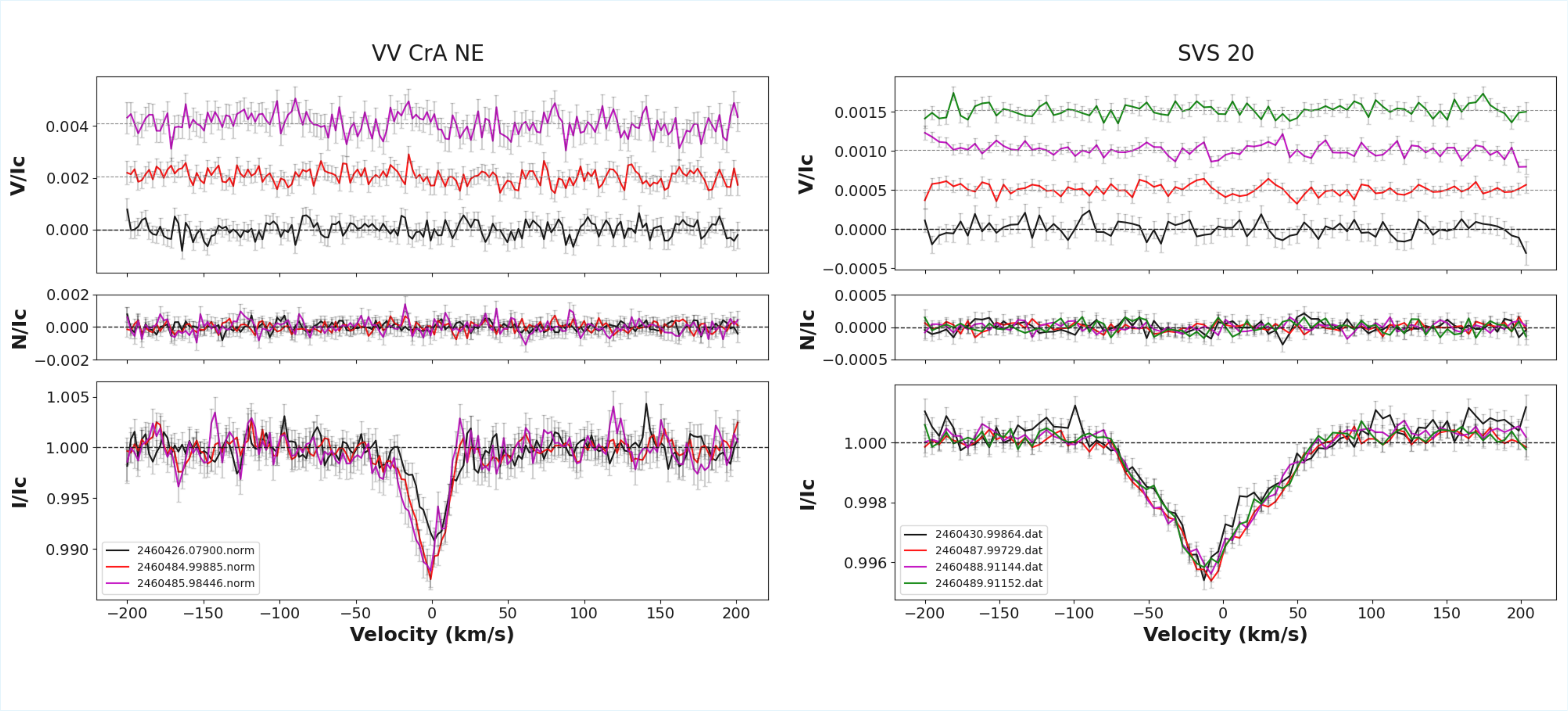}
\caption{\label{fig:lsd}LSD profiles for each star. The different colours indicate the different observations. The Y-axes are $I$/$Ic$, $N$/$Ic$, and $V$/$Ic$, where $Ic$ is the intensity of the stellar continuum. The $I$ profile represents the total intensity, the Stokes $V$ profile corresponds to the circular polarisation, and the Stokes $N$ profile corresponds to the null spectra. A pixel size of 4.8 instead of 2.4 is applied for stars with \vsini\ > 35 km s\(^{-1}\).}
\end{figure*}

\begin{table*}[h]
\caption{LSD Stokes $I$ fitting results, veiling values, and longitudinal magnetic field detection and measurements.}
\label{tab:vals}\centering
\centering
{\fontsize{9}{11}\selectfont
\begin{tabular}{l c c c c c c c c r@{\,$\pm$\,}l}
\hline
\hline
ID & BJD & Depth $I$ & \vsini & \vrad & $FAP$ & $B$ & $r_{\rm H}$ & $r_{\rm K}$ & \multicolumn{2}{c}{\bl} \\
     & -2\,400\,000 & $\times 10^{-4}$ & (\kms) & (\kms) & & detection & & & \multicolumn{2}{c}{(G)}   \\
\hline
\textbf{V806 Tau*} & 59900.12063 & $210 \pm 1$ & $27.4 \pm 0.5$ & $12.7 \pm 0.3$ & $7.1\times10^{-6}$ & DD & $0.7 \pm 0.1$ & $2 \pm 0.1$ & {\bf80} & {\bf18} \\
         & 59901.12629 & $160 \pm 1$ & $25.7 \pm 0.6$ & $10.9 \pm 0.4$ & $5.5\times10^{-6}$ & DD & $1.2 \pm 0.1$ & $3.3 \pm 0.2$ & {\bf77} & {\bf21} \\
         & 59915.02084 & $140 \pm 1$ & $25.6 \pm 0.8$ & $12.7 \pm 0.5$ & $1.9\times10^{-3}$ & ND & $1.6 \pm 0.1$ & -- & 100 & 25 \\
\hline
\textbf{MHO 3*}    & 59900.09753 & $100 \pm 1$ & $56.1 \pm 0.8$ & $13.1 \pm 0.6$ & $1.2\times10^{-1}$ & ND & $1 \pm 0.1$ & $2 \pm 0.1$ & -92 & 50 \\
         & 59914.97152 & $100 \pm 1$ & $53.9 \pm 0.7$ & $13.5 \pm 0.4$ & $1.9\times10^{-1}$ & ND & $0.5 \pm 0.1$ & $1.8 \pm 0.1$ & -142 & 42 \\
         & 59918.01608 & $60 \pm 1$ & $50.5 \pm 1.2$ & $12.9 \pm 0.9$ & $9.1\times10^{-1}$ & ND & -- & -- & -85 & 67 \\
         & 59945.96152 & $100 \pm 1$ & $55.4 \pm 0.7$ & $14.2 \pm 0.4$ & $7.9\times10^{-6}$ & DD & $0.4 \pm 0.1$ & $1.3 \pm 0.1$ & {\bf-144} &  {\bf45} \\
\hline
\textbf{MHO 2*}    & 59918.04028 & $100 \pm 1$ & $30.4 \pm 1.1$ & $13.5 \pm 0.7$ & $7.7\times10^{-1}$ & ND & $0.3 \pm 0.1$ & $0.6 \pm 0.1$ & -69 & 29 \\
         & 59946.87428 & $200 \pm 1$ & $30.5 \pm 0.9$ & $13.9 \pm 0.5$ & $2.5\times10^{-10}$ & DD & $0.3 \pm 0.1$ & $0.6 \pm 0.1$ & {\bf-97} & {\bf17} \\
         & 59947.85486 & $160 \pm 1$ & $31.5 \pm 1.0$ & $12.7 \pm 0.6$ & $2.8\times10^{-5}$ & MD & $0.5 \pm 0.1$ & $0.6 \pm 0.1$ & 26 & 19 \\ 
\hline
IRAS 04292+2422 W & 59946.90951 & $900 \pm 20$ & $7.9 \pm 0.2$ & $28.1 \pm 0.3$ & $2.2\times10^{-1}$ & ND & $0.1 \pm 0.1$ & $0.2 \pm 0.1$ & -2.4 & 2.8 \\
                  & 59947.89613 & $1010 \pm 20$ & $7.8 \pm 0.3$ & $28.1 \pm 0.3$ & $9.6\times10^{-1}$ & ND & $0.2 \pm 0.1$ & $0.2 \pm 0.1$ & -1.4 & 2.7 \\
                  & 59952.97509 & $900 \pm 30$ & $7.9 \pm 0.2$ & $28.1 \pm 0.3$ & $3.5\times10^{-1}$  & ND & $0.2 \pm 0.1$ & $0.2 \pm 0.1$ & -3.1 & 3.1 \\
\hline
IRAS 04113+2758 N* & 59948.93997 & $160 \pm 10$ & $17.9 \pm 0.8$ & $12.7 \pm 0.4$ & $1.2\times10^{-1}$ & ND & $1.1 \pm 0.1$ & $2.4 \pm 0.1$ & 11 & 10 \\
                  & 59953.85844 & $140 \pm 10$ & $17.2 \pm 1.0$ & $12.4 \pm 0.5$ & $1.6\times10^{-2}$ & ND & $2 \pm 0.1$ & $3.1 \pm 0.1$ & -4 & 12 \\
                  & 59954.83884 & $140 \pm 10$ & $18.3 \pm 1.2$ & $12.4 \pm 0.6$ & $4.7\times10^{-1}$ & ND & $1.6 \pm 0.1$ & $2.6 \pm 0.1$ & -7 & 12 \\
\hline
IRAS 04369+2539* & 59903.13177 & $80 \pm 1$ & $40.6 \pm 0.5$ & $13.8 \pm 0.3$ & $6.4\times10^{-1}$ & ND & $1.6 \pm 0.1$ & $4.7 \pm 0.6$ & 9 & 52\\
                & 59905.05519 & $90 \pm 1$ & $40.7 \pm 0.6$ & $14.5 \pm 0.4$ & $8.1\times10^{-1}$ & ND & $1.4 \pm 0.1$ & $5.2 \pm 0.8$ & -10 & 62 \\
                & 59915.06549 & $90 \pm 1$ & $41.9 \pm 0.7$ & $12.7 \pm 0.5$ & $4.9\times10^{-1}$ & ND & $1.5 \pm 0.1$ & $2.3 \pm 0.2$ & 42 & 44 \\
\hline
HL Tau & 58384.11335 & $60 \pm 1$ & $52.5 \pm 2.5$ & $11.9 \pm 2.5$ & $9.1\times10^{-1}$ & ND & $1.2 \pm 0.1$ & $4.5 \pm 0.2$ & 31 & 71 \\
       & 58385.09837 & $70 \pm 1$ & $51.9 \pm 1.7$ & $12.8 \pm 1.2$ & $1.5\times10^{-1}$ & ND & $1.1 \pm 0.1$ & $6.4 \pm 0.3$ & -92 & 69 \\
       & 58387.08102 & $70 \pm 1$ & $49.4 \pm 1.6$ & $13.4 \pm 1.1$ & $1.5\times10^{-1}$ & ND & $1.4 \pm 0.1$ & $8.6 \pm 0.4$ & -125 & 79 \\
       & 58499.80432 & $60 \pm 1$ & $57.7 \pm 3.7$ & $12.9 \pm 2.9$ & $4.4\times10^{-1}$ & ND & $1.5 \pm 0.1$ & $8.8 \pm 0.4$ & -25 & 78 \\
       & 58500.79189 & $60 \pm 1$ & $54.2 \pm 2.0$ & $13.7 \pm 1.5$ & $7.5\times10^{-1}$ & ND & $1.5 \pm 0.2$ & $12 \pm 0.6$ & -58 & 69 \\
       & 59111.08910 & $60 \pm 1$ & $51.1 \pm 1.6$ & $12.1 \pm 1.2$ & $1.2\times10^{-1}$ & ND & $1.7 \pm 0.1$ & $6.9 \pm 0.3$ & -13 & 60 \\
       & 59115.05352 & $60 \pm 1$ & $52.9 \pm 2.1$ & $14.8 \pm 1.5$ & $5.4\times10^{-1}$ & ND & $1.8 \pm 0.1$ & $9.7 \pm 0.5$ & 54 & 58 \\
\hline
\textbf{IRAS 16191-1936*} & 60388.10393 & $80 \pm 1$ & $59.5 \pm 0.7$ & $-5.2 \pm 0.4$ & $5.3\times10^{-1}$ & ND & $0.9 \pm 0.1$ & $2.9 \pm 0.2$ & 38 & 56 \\
                & 60433.02290 & $100 \pm 1$ & $53.9 \pm 0.5$ & $-7.7 \pm 0.9$ & $4.8\times10^{-13}$ & DD & $0.4 \pm 0.1$ & $1.7 \pm 0.1$ & {\bf204} & {\bf46} \\
                & 60487.76569 & $80 \pm 1$ & $53.0 \pm 0.3$ & $-9.1 \pm 0.9$ & $9.1\times10^{-8}$ & DD & $0.6 \pm 0.1$ & $2.3 \pm 0.1$ & {\bf151} & {\bf44} \\
\hline
\textbf{L1689 SNO2*} & 58618.10070 & $150 \pm 1$ & $24.1 \pm 0.7$ & $-7.6 \pm 0.4$ & $1.4\times10^{-1}$ & ND & $0.6 \pm 0.1$ & $2.1 \pm 0.1$& 56 & 44 \\ 
           & 60509.89855 & $170 \pm 1$ & $23.3\pm 0.5$ & $-8.1 \pm 0.3$ & $4.1\times10^{-13}$  & DD & $0.6 \pm 0.1$ & $1.9 \pm 0.1$ & {\bf-130} & {\bf21} \\
           & 60517.85402 & $160 \pm 1$ & $23.4 \pm 0.5$ & $-7.0 \pm 0.3$ & $< 10^{-15}$ & DD & $0.6 \pm 0.1$ & $2.2 \pm 0.1$ & {\bf-180} & {\bf24} \\
\hline
IRAS 03247+3001* & 59946.84774 & $90 \pm 1$ & $67.2 \pm 1.2$ & $12.0 \pm 0.8$ & $4.1\times10^{-1}$ & ND & $0.7 \pm 0.1$ & $2.5 \pm 0.1$ & -260 & 133 \\ 
                & 59947.81412 & $100 \pm 1$ & $67.4 \pm 1.9$ & $11.8 \pm 1.4$ & $9.9\times10^{-1}$ & ND & $1 \pm 0.2$ & $3.8 \pm 0.2$ & 22 & 123 \\
                & 59948.84853 & $120 \pm 1$ & $68.2 \pm 2.7$ & $-6.4 \pm 1.4$ & $9.0\times10^{-1}$ & ND & $1.4 \pm 0.3$ & $3 \pm 0.2$ & 142 & 99 \\
                & 59953.90247 & $100 \pm 1$ & $63.5 \pm 1.3$ & $-0.8 \pm 1.3$ & $5.4\times10^{-1}$ & ND & $0.7 \pm 0.1$ & $2.6 \pm 0.2$ & -12 & 72 \\
                & 59954.88269 & $110 \pm 1$ & $66.9 \pm 1.5$ & $-2.5 \pm 1.3$ & $7.4\times10^{-1}$ & ND & $0.9 \pm 0.1$ & $3.1 \pm 0.2$ & 82 & 83 \\
                & 59955.89845 & $90 \pm 1$ & $76.9 \pm 4.1$ & $17.3 \pm 3.0$ & $8.3\times10^{-1}$ & ND & $0.9 \pm 0.1$ & $2.9 \pm 0.2$ & -42 & 83 \\
\hline
V512 Per & 59090.03171 & $140 \pm 10$ & $23.8 \pm 2.7$ & $13.9 \pm 1.1$ & $9.6\times10^{-1}$ & ND & $0.9 \pm 0.1$ & $0.8 \pm 0.2$ & 55 & 37 \\
& 59948.87769 & $140 \pm 10$ & $22.4 \pm 2.9$ & $14.6 \pm 1.8$ & $9.6\times10^{-1}$ & ND & $2.1 \pm 0.1$ & $2.8 \pm 0.3$ & 28 & 55 
          \\ 
\hline
VV CrA SW & 60402.15170 & $90 \pm 1$ & $36.8 \pm 2.2$ & $4.6 \pm 1.2$ & $3.6\times10^{-2}$  & ND & -- & -- & 141 & 56 \\
          & 60425.12513 & $90 \pm 1$ & $31.5 \pm 1.3$ & $2.3 \pm 0.6$ & $9.6\times10^{-2}$ & ND & -- & -- & -107 & 74 \\
          & 60484.02208 & $40 \pm 1$ & $43.6 \pm 3.2$ & $4.1 \pm 1.5$ & $3.1\times10^{-1}$ & ND & -- & -- & 199 & 159 \\
          & 60484.96903 & $40 \pm 1$ & $31.8 \pm 1.7$ & $7.2 \pm 1.6$ & $3.1\times10^{-1}$ & ND & -- & -- & -18 & 157 \\
\hline
VV CrA NE* & 60426.07900 & $80 \pm 1$ & $13.8 \pm 2.5$ & $-0.4 \pm 0.5$ & $6.3\times10^{-1}$ & ND & -- & -- & 17 & 72 \\
          & 60484.99885 & $110 \pm 1$ & $15.2 \pm 1.4$ & $-3.1 \pm 0.2$ & $4.5\times10^{-1}$ & ND & $4.6 \pm 0.3$ & $8.8 \pm 0.4$ & 87 & 49 \\
          & 60485.98446 & $110 \pm 1$ & $16.6 \pm 2.3$ & $-6.2 \pm 0.6$ & $7.2\times10^{-1}$ & ND & $7.9 \pm 0.4$ & $11.6 \pm 0.5$ & 169 & 74 \\
\hline
SVS 20 S* & 60430.99864 & $30 \pm 1$ & $61.6 \pm 2.2$ & $-10.6 \pm 1.3$ & $7.5\times10^{-1}$ & ND & $2 \pm 0.2$ & $10.9 \pm 0.5$ & 57 & 104 \\ 
         & 60487.99729 & $30 \pm 1$ & $59.7 \pm 1.0$ & $-9.8 \pm 0.8$ & $1.2\times10^{-2}$ & ND & $1.8 \pm 0.1$ & $18 \pm 0.9$ & 120 & 68 \\
         & 60488.91144 & $30 \pm 1$ & $58.6 \pm 1.1$ & $-10.9 \pm 0.7$ & $2.3\times10^{-2}$ & ND & $2.2 \pm 0.2$ & $9.2 \pm 0.6$ & -49 & 61 \\
         & 60489.91152 & $30 \pm 1$ & $59.5 \pm 1.2$ & $-9.5 \pm 0.8$ & $7.5\times10^{-1}$ & ND & $1.8 \pm 0.2$ & $15.2 \pm 0.8$ & 170 & 82 \\
\hline
\end{tabular}}
\tablefoot{From left to right the columns show the depth, the projected rotational velocity (\vsini), the radial velocity (\vrad), computed with a physical stellar model (see Appendix \ref{appendix:upperlimits}), the false alarm probability, the magnetic detection status (defined as DD for definite detection, MD for marginal detection, and ND for no detection), the veiling values in H and K band, the longitudinal magnetic field values, and their corresponding uncertainties. The stars indicated in bold are detected magnetic stars (at least one DD in the different observations). Sources with an * symbol are flat-spectrum targets.}
\end{table*}

\end{document}